\documentclass[iop,apj]{emulateapj}
\usepackage{apjfonts}
\usepackage{graphicx}

\begin{document}

\title{The {\it Hubble Space Telescope} UV Legacy Survey of 
Galactic Globular Clusters.\\ 
VII. Implications from the Nearly Universal Nature of 
Horizontal Branch Discontinuities\altaffilmark{1}}
\shorttitle{Horizontal Branch Discontinuities}

\author{
T.M. Brown\altaffilmark{2}, 
S. Cassisi\altaffilmark{3},
F. D'Antona\altaffilmark{4},
M. Salaris\altaffilmark{5},
A.P. Milone\altaffilmark{6},
E. Dalessandro\altaffilmark{7},
G. Piotto\altaffilmark{8,9},\\
A. Renzini\altaffilmark{9},
A.V. Sweigart\altaffilmark{10},
A. Bellini\altaffilmark{2}, 
S. Ortolani\altaffilmark{8,9},
A. Sarajedini\altaffilmark{11},\\ 
A. Aparicio\altaffilmark{12,13},
L.R. Bedin\altaffilmark{9},
J. Anderson\altaffilmark{2},
A. Pietrinferni\altaffilmark{3},
and 
D. Nardiello\altaffilmark{8,9}
}

\altaffiltext{1}{Based on observations made with the NASA/ESA {\it Hubble
Space Telescope}, obtained at the Space Telescope Science Institute, which
is operated by the Association of Universities for Research in Astronomy, 
Inc., under NASA contract NAS 5-26555.  These observations are associated
with program GO-13297.}

\altaffiltext{2}{
Space Telescope Science Institute, 3700 San Martin Drive,
Baltimore, MD 21218, USA;  tbrown@stsci.edu; jayander@stsci; bellini@stsci.edu
}

\altaffiltext{3}{
INAF-Osservatorio Astronomico di Teramo, Via Mentore
Maggini s.n.c., I-64100 Teramo, Italy; cassisi@oa-teramo.inaf.it;
pietrinferni@oa-teramo.inaf.it
}

\altaffiltext{4}{
INAF-Osservatorio Astronomico di Roma, Via Frascati
33, I-00040 Monteporzio Catone, Roma, Italy;
dantona@oa-roma.inaf.it
}

\altaffiltext{5}{
Astrophysics Research Institute, Liverpool John Moores University,
Liverpool Science Park, IC2 Building, 146 Brownlow Hill, Liverpool L3 5RF,
UK; M.Salaris@ljmu.ac.uk}

\altaffiltext{6}{
Research School of Astronomy and Astrophysics, The
  Australian National University, Cotter Road, Weston, ACT, 2611,
  Australia; milone@mso.anu.edu.au
}

\altaffiltext{7}{
Dipartimento di Fisica e Astronomia, Universit\`a
degli Studi di Bologna, Viale Berti Pichat 6/2, I-40127 Bologna,
Italy; emanuele.dalessandr2@unibo.it
}

\altaffiltext{8}{
Dipartimento di Fisica e Astronomia ``Galileo
Galilei'', Universit\`a di Padova, Vicolo dell'Osservatorio 3,
I-35122 Padova, Italy; giampaolo.piotto@unipd.it; sergio.ortolani@unipd.it;
domenico.nardiello@unipd.it}

\altaffiltext{9}{
INAF-Osservatorio Astronomico di Padova, Vicolo
dell'Osservatorio 5, I-35122 Padova, Italy
alvio.renzini@oapd.inaf.it; luigi.bedin@oapd.inaf.it
}

\altaffiltext{10}{
NASA Goddard Space Flight Center, Greenbelt, MD 20771, USA;
allen.sweigart@gmail.com}

\altaffiltext{11}{
Department of Astronomy, University of Florida, 211
Bryant Space Science Center, Gainesville, FL 32611, USA;
ata@astro.ufl.edu
}

\altaffiltext{12}{
Instituto de Astrof\'\i sica de Canarias. 
Calle V\'\i a L\'actea s/n. E38200 - La Laguna, 
Tenerife, Canary Islands, Spain. aaj@iac.es
}

\altaffiltext{13}{
University of La Laguna. Avda. Astrof\'isico Fco. S\'anchez, s/n. E38206, 
La Laguna, Tenerife, Canary Islands, Spain}

\submitted{Accepted for publication in The Astrophysical Journal}

\begin{abstract}

The UV-initiative {\it Hubble Space Telescope} Treasury survey of
Galactic globular clusters provides a new window into the phenomena
that shape the morphological features of the horizontal branch (HB).
Using this large and homogeneous catalog of UV and blue photometry, we
demonstrate that the HB exhibits discontinuities that are remarkably
consistent in color (effective temperature).  This consistency is apparent
even among some of the most massive clusters hosting multiple distinct
sub-populations (such as NGC~2808, $\omega$~Cen, and NGC~6715),
demonstrating that these phenomena are primarily driven by atmospheric
physics that is independent of the underlying population properties.
However, inconsistencies arise in the metal-rich clusters NGC~6388 and
NGC~6441, where the discontinuity within the blue HB (BHB)
distribution shifts 
$\sim$1,000~K to 2,000~K hotter.  We
demonstrate that this shift is likely due to a large helium
enhancement in the BHB stars of these clusters, which in turn affects
the surface convection and evolution of such stars.  Our survey also
increases the number of Galactic globular clusters known to host
blue-hook stars (also known as late hot flashers) from 6 to 23
clusters.  These clusters are biased toward the bright end of the
globular cluster luminosity function, confirming that blue-hook stars
tend to form in the most massive clusters with significant
self-enrichment.

\end{abstract}

\keywords{globular clusters: general -- stars: atmospheres -- stars: evolution
-- stars: horizontal branch -- ultraviolet: stars}

\section{Introduction}

Although globular clusters represent the best available laboratories
for constraining stellar evolution models,
we now know they are not simple stellar populations.  Evidence for complex
populations are
manifested in all phases of stellar evolution.  On the main sequence
(MS) and red giant branch (RGB), high-precision photometry reveals
distinct sequences that are most prominent in massive clusters such as
$\omega$~Cen (e.g., Anderson 1997; Bedin et al.\ 2004; Ferraro et
al.\ 2004), NGC~2808 (e.g., D'Antona et al.\ 2005; Piotto et al 2007;
Milone et al.\ 2015b), M2 (Milone et al.\ 2015a),
and NGC~6715 (e.g., Layden \& Sarajedini 2000; Siegel et al.\ 2007;
Piotto et al.\ 2012), but the phenomenon is apparently universal
(Piotto et al.\ 2012, 2015).  The formation mechanisms for these
multiple populations remain unclear (see Renzini et al.\ 2015).

Even before the existence of stellar sub-populations in globular clusters
was known, there was considerable evidence for peculiarities
in the morphology of the horizontal branch (HB).  In particular, there
was the well-known 
``second parameter'' problem, first mentioned by Sandage \&
Wallerstein (1960; see also Sandage \& Wildey 1967; van den Bergh 1967);
it refers to the observation that parameters other
than metallicity (such as age and He abundance) must affect the
color distribution of HB stars (see Catelan 2009 for a review).  
In those clusters where the HB distribution is sufficiently broad in
color, a series of discontinuities is also evident, although the
appearance of these features varies with the bandpasses employed,
manifesting themselves as gaps, jumps, overluminous stars, or subluminous
stars.  Three prominent examples of such discontinuities are
the ``Grundahl jump'' (G-jump) within the blue HB (BHB) at
$\sim$11,500~K (Grundahl et al.\ 1998, 1999), the ``Momany jump'' (M-jump)
within the extreme HB (EHB) at $\sim$20,000~K (Momany et al.\ 2002,
2004), and the gap between the EHB and ``blue-hook'' stars (also known
as ``late hot flasher'' stars; D'Cruz et al.\ 1996, 2000), spanning
$\sim$32,000--36,000~K (Sweigart 1997; Brown et al.\ 2001).  
Here we have adopted the usual naming convention
for HB stars: EHB stars are those at $T_{\rm eff} \gtrsim$20,000~K,
while BHB stars are those hotter than the RR~Lyrae instability strip 
(i.e., hotter than $\sim$8,000~K) but cooler than the EHB.
Both Grundahl et al.\ (1999) and Momany et al.\ (2004) noted that their
respective jumps appear to be ubiquitous features of globular clusters
hosting sufficient numbers of BHB and EHB stars, but the surprising
consistency of HB gaps in distinct clusters was recognized somewhat
earlier (Ferraro et al.\ 1998).  
Similarly, the blue-hook phenomenon
appears to be common in those clusters hosting an HB population
that extends sufficiently far to the blue, but because optical colors
become degenerate at the temperatures of EHB stars, UV photometry is
needed to confirm their presence (D'Cruz et al.\ 2000; Brown et
al.\ 2001; Dalessandro et al.\ 2008; Dieball et al.\ 2009; Brown et al.\ 2010).

In a review of hot stars in globular clusters, Moehler (2001) explored
various explanations for HB discontinuities, including diverging
evolutionary paths, mass loss, distinctions in CNO or rotation rates,
dynamical interactions, atmospheric processes, He mixing in red
giants, and statistical fluctuations.  With time, it has become
increasingly clear that atmospheric processes play a dominant role in
these HB features.  Spectroscopy of stars on either side of the
hottest discontinuity demonstrates that, compared to normal EHB stars,
blue-hook stars have atmospheres greatly enhanced in He and C (Moehler
et al.\ 2011; Brown et al.\ 2012), as expected if they formed through
a late He core flash that mixed the 
H-rich envelope with the inner convective regions
(a process known as flash mixing; Sweigart 1997; Cassisi et al.\ 2003).  
The BHB stars hotter than the G-jump
exhibit metal abundances enhanced via radiative levitation,
He abundances diminished via gravitational settling (Moehler et
al.\ 1999, 2000; Behr 2003; Pace et al.\ 2006), and lower surface
gravities than expected from canonical BHB models (see Moehler 2001
and references therein).  Stellar evolution models that include atomic
diffusion have had qualitative
success reproducing the observed abundance patterns
in BHB stars (Michaud et al.\ 2007, 2008).  Sweigart (2002) first
noted that the onset of radiative levitation on the BHB coincided with
the disappearance of surface convection.  The interplay between atomic
diffusion and surface convection was explored more fully by Cassisi \&
Salaris (2013); they noted that by itself, surface convection should
not be enough to suppress radiative levitation in stars cooler than
the G-jump, implying that other processes, such as turbulence and
rotation, must also play a role.

In addition to abundance differences, there is a clear bimodality in
rotation on the BHB (Recio-Blanco et al.\ 2002, 2004; Behr 2003);
stars hotter than the G-jump are generally slow rotators ($v$~sin~$i <
8$~km s$^{-1}$), while stars cooler than the G-jump are generally
fast rotators (with $v$~sin~$i$ as high as 40 km s$^{-1}$).  Although
the source of this bimodality remains unclear, Recio-Blanco et
al.\ (2002, 2004) have speculated that the dearth of fast rotators
hotter than the G-jump could be due to a loss of angular momentum
through stellar winds, driven by the high atmospheric metallicity at
such temperatures.  Quievy et al.\ (2009) have argued that meridional
circulation in the fast rotators plays a role in the disruption of
atomic diffusion in stars cooler than the G-jump.

The consistency of these HB features, and any exceptions to that
consistency, are difficult to explore with the heterogeneous
observations available in the literature.  However, a new catalog of
UV and blue photometry, resulting from a {\it Hubble Space Telescope
  (HST)} Treasury survey of globular clusters, is well suited to this
task (Piotto et al.\ 2015).  In this paper, the seventh in a series
associated with the survey, we characterize these HB
features in a diverse set of 53 clusters, including clusters with
significantly complex populations.  We then use these comparisons to
explore the implications for the atmospheric phenomena and the
evolutionary history of hot stars in globular clusters. 

\section{Data}

Our analysis employs photometry obtained with the UVIS channel of the
Wide Field Camera 3 (WFC3) on board {\it HST}, largely derived from
the UV-initiative Treasury program 13297 (PI: Piotto), but
supplemented with data from other Guest Observer programs (e.g.,
GO-12311, GO-12605; PI Piotto).  
For $\omega$~Cen, we use
the photometry of Bellini et al. (2013a, and also in prep.), which
draws upon archival WFC3 data, including those from calibration programs.  The
observations employed the F275W (near-UV), F336W ($U$), and F438W
($B$) filters (Figure~1).
The Treasury program and data reduction are fully
described in Piotto et al.\ (2015), and the zero-point calibration and
differential reddening corrections are detailed in Milone et
al.\ (2015a). Therefore, we will only briefly summarize the
observations and data reduction here.

%fig1
\begin{figure*}[t]
\begin{center}
\includegraphics[width=6.5in]{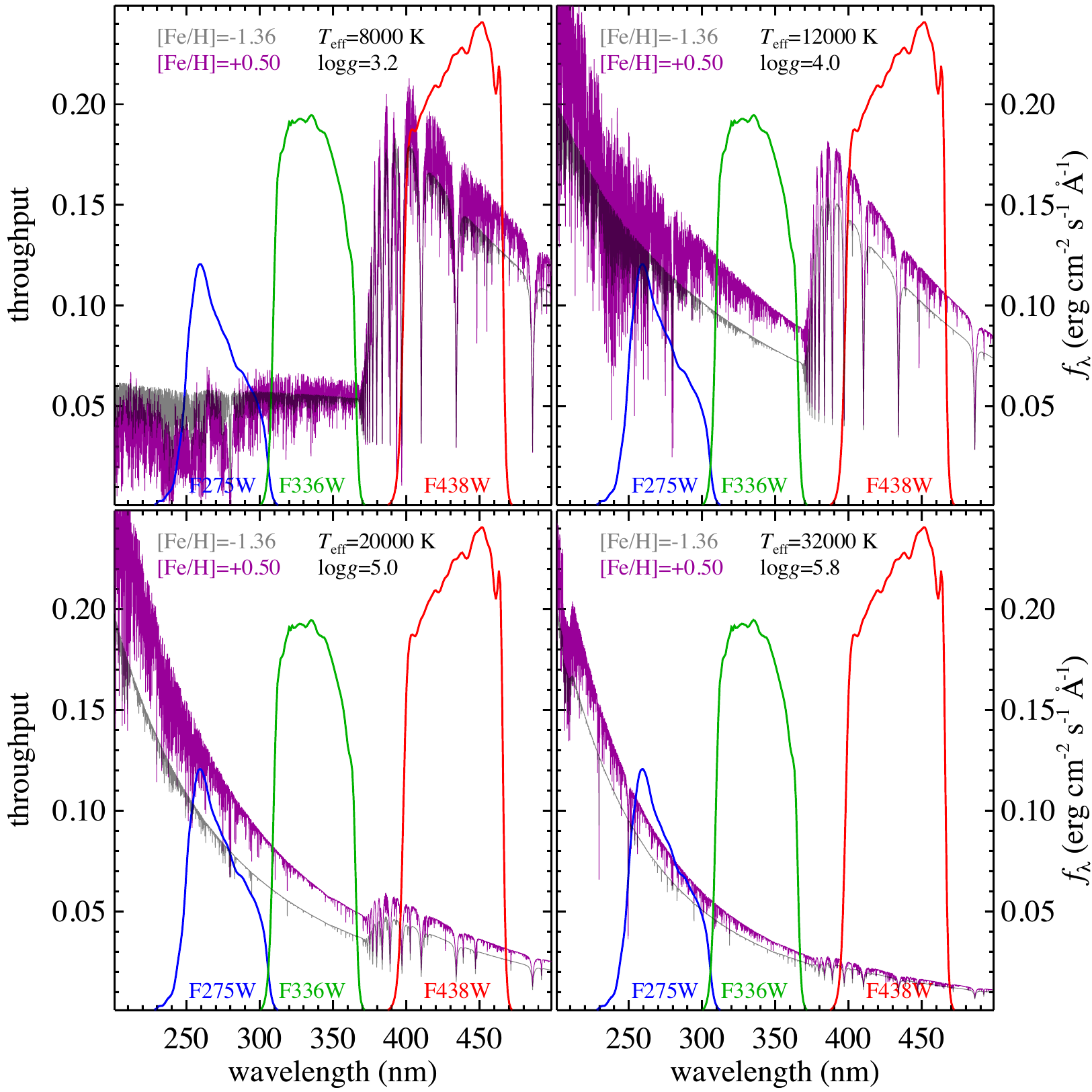}
\end{center}
\caption{
The arbitrarily normalized 
spectra of four stars along the HB ($T_{\rm eff}$ and surface gravity 
labeled), at the NGC~2808 metallicity ({\it grey curve}), compared to 
the spectra when the metals are enhanced to three times the solar value
({\it purple curve}).  Although we show the effects of metallicity enhancement
at each temperature, such enhancements (associated with radiative levitation
in the atmosphere) are only observed at temperatures
hotter than the G-jump (see text).
For comparison, we show the WFC3 bandpasses
employed in our analysis: F275W ({\it blue curve}), F336W ({\it green
curve}), and F438W ({\it red curve}).  The 
$(m_{\rm F275W} - m_{\rm F336W})$ color tracks absorption from Fe line
blanketing in the near-UV, while the $(m_{\rm F336W} - m_{\rm F438W})$
color spans the Balmer discontinuity.  The $C_{\rm
  F275W,F336W,F438W}$ index combines both colors, and is sensitive to both
radiative levitation and surface gravity in BHB stars. The high-resolution
spectra shown here were calculated using the ATLAS9 and SYNTHE codes
(Kurucz 1993; Sbordone et al.\ 2004), for consistency
with the Castelli \& Kurucz (2003) spectra used in our analysis.}
\end{figure*}

Images from program 13297 were obtained from Aug 2013 to Jun 2015.  The
data were corrected for charge transfer inefficiency using a
pixel-based algorithm developed by Anderson \& Bedin (2010) for use
with the Advanced Camera for Surveys, but later modified for use
with the WFC3.  Photometry was performed on each individual exposure
using a library of spatially-variable empirical point spread
functions, combined into a single measurement for each star, and
placed in the VEGAMAG system.

%fig2
\begin{figure*}[t]
\begin{center}
\includegraphics[width=6.0in]{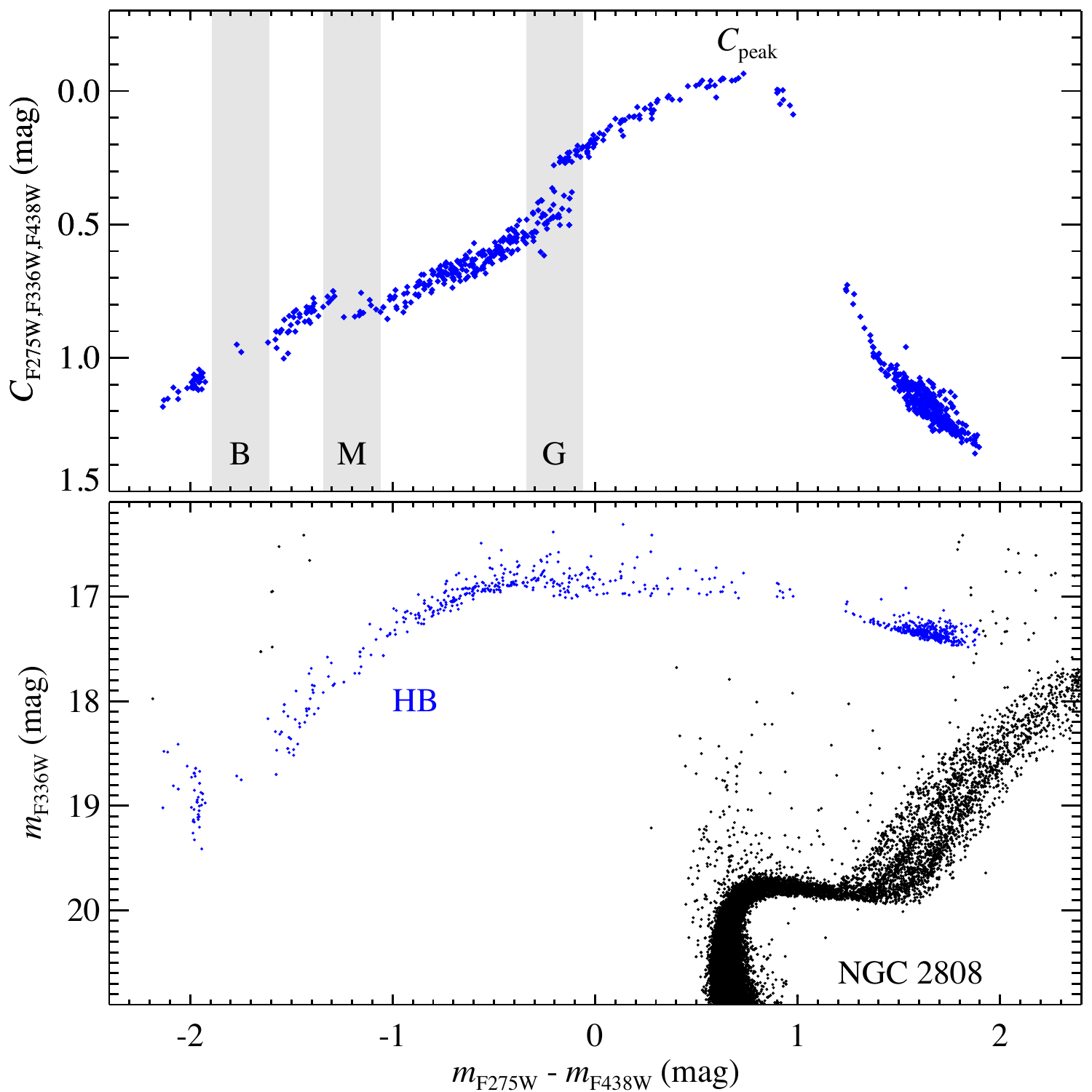}
\end{center}
\caption{{\it Bottom panel:} The CMD of NGC~2808 ({\it points}), highlighting
the HB distribution ({\it blue points, labeled}).  {\it Top panel:}
The same HB distribution, but the ordinate has been replaced by the
$C_{F275W,F336W,F438W}$ color index.  In this CCP, the various HB
discontinuities are easily distinguished: the
Grundahl jump (G), the Momany jump (M), and the gap between the blue-hook
and normal EHB stars (B). The large gap to the red of the Grundahl
jump is the RR~Lyrae instability strip.  We use the peak in the
$C_{\rm F275W,F336W,F438W}$ index ({\it labeled}) to align HB distributions
to each other and to models.
}
\end{figure*}

The primary goal of these imaging programs was to study the multiple
populations on the MS and RGB.  Using observations of NGC~6752,
Milone et al.\ (2013) demonstrated that the multiplicity of
globular cluster populations could be distinguished using the color index 
$C_{\rm F275W,F336W,F438W} = 
(m_{\rm F275W} - m_{\rm F336W}) - (m_{\rm F336W} - m_{\rm  F438W})$;
Piotto et al.\ (2015) subsequently demonstrated the utility of this index 
in a large set of clusters.
For the cool stars on the MS and RGB, the F275W filter
includes an OH molecular band, the F336W filter includes an NH
molecular band, and the F438W filter includes CN and CH
bands (Milone et al.\ 2012).  We will not explore these
features here, but they also affect the morphology of the red clump,
which falls at similarly cool temperatures.

Although the program was driven by a desire to characterize the MS and RGB,
the same UV data provide insight into hot stars in later evolutionary phases.
The HB population is well
exposed, with photometric errors of $\sim$0.01~mag and nearly 100\%
completeness.  As on the MS and RGB,
the $C_{\rm F275W,F336W,F438W}$ color index is useful for characterizing
the HB morphology, but for entirely
different reasons (Figure~1).  At hotter temperatures, the
$(m_{\rm F275W} - m_{\rm F336W})$ color tracks absorption from Fe line
blanketing in the near-UV, while the $(m_{\rm F336W} - m_{\rm F438W})$
color spans the Balmer discontinuity, making the
$C_{\rm F275W,F336W,F438W}$ index sensitive
to atmospheric metallicity and surface gravity in HB stars.  
Note that these broad-band filters are relatively insensitive to 
atmospheric He abundance, over the range generally encountered on the HB.  
Specifically, taking the HB spectra at each of the four temperatures shown 
in Figure~1, if we decrease the atmospheric $Y$ to 0.01 (simulating the effects
of He gravitational settling), or if we increase the atmospheric $Y$ to 0.40
(as associated with a self-enriched sub-population), the variations
within each of our bands are $\leq$0.03~mag, and generally $\sim$0.01~mag
(see also Sbordone et al.\ 2011; Girardi et al.\ 2007; 
Dalessandro et al.\ 2013).

In Figure~2, we show the  
$C_{\rm F275W,F336W,F438W}$ index alongside the
color-magnitude diagram (CMD) of NGC~2808, 
a massive globular cluster long known to host
a peculiar HB morphology (Sosin et al.\ 1997; Bedin et al.\ 2000).  
We highlight three discontinuities in the HB distribution: the
G-jump (labeled ``G''), the M-jump (labeled ``M''), and
the gap between the blue-hook stars and the normal EHB (labeled
``B'').  While these breaks can be discerned in the individual colors
comprising the $C_{\rm F275W,F336W,F438W}$ index, they are amplified when
the colors are combined in this index.  Furthermore, in these crowded fields,
the photometric errors are correlated amongst the various bands; 
in the color-color plane (CCP), these errors are suppressed in both axes,
while a CMD will suppress them in only one axis.  
In addition to these three prominent features, there are small gaps 
(e.g., at 0.4~mag and 0.8~mag in $m_{\rm F275W} - m_{\rm F438W}$ color),
but these do not correspond to features within the HB
distributions of the other clusters we will consider here, and could
be statistical fluctuations (see, e.g., Catelan et al.\ 1998).
For reference, the peak in the
$C_{\rm F275W,F336W,F438W}$ index corresponds to an effective temperature of
$\sim$8,600~K, driven primarily by the sensitivity to the Balmer
discontinuity in the ($m_{\rm F336W} - m_{\rm F438W}$) component of the
index.  This peak is well-defined in the CCP, falls in the 
middle of the HB color range, and is populated in nearly all of the clusters
in our sample, so we shall use it as a reference point, hereafter
called $C_{\rm peak}$.

\section{Zero-Age Horizontal Branch Models}

The ZAHB models of Brown et al.\ (2010) that are used in the present
paper were computed with a highly-modified version of the original
Princeton stellar evolution code (Schwarzschild \& H\"arm 1965). This
code has been extensively updated over the years
(Sweigart \& Demarque 1972; Sweigart \& Gross 1974, 1978; 
Sweigart 1987, 1997). The equation of state (EOS) is
based on a tabulation of the Fermi-Dirac integrals and the various
thermodynamic functions for both the non-relativistic and relativistic
regimes. At low temperatures, the EOS solves the Saha equation for the
ionization of H and He, as well as the first ionization of 10 heavy
elements, plus the formation of H$_2$. The code also incorporates the
OPAL radiative opacities of Rogers \& Iglesias (1992), while at low
temperatures, the molecular opacities of Bell (1995, private
communication) are used. The nuclear reaction rates are taken from
Caughlan \& Fowler (1988), except for the $^{12}$C($\alpha$,
$\gamma$)$^{16}$O reaction, where the higher rate suggested by Weaver \&
Woosley (1993) has been adopted. A further discussion of the input
physics in our evolution code can be found in 
the description of the related PGPUC code of Valcarce et al.\ (2012).

This code has also been highly automated and, as a result, can follow
the evolution of a globular-cluster star continuously from the MS up
the RGB and then through the He-core flash, HB, and asymptotic-giant
branch phases in a single computer run. For example, all of the HB
sequences in Brown et al. (2010) were evolved continuously through the
He flash, thereby avoiding the need to construct separate ZAHB
models. Convective overshooting and semi-convection in our HB models
have been treated according to the method of Robertson \& Faulkner
(1972). This method is applied between iterations of the Henyey method
and ensures that the radiative and adiabatic gradients agree to better
than 10$^{-4}$ at the convective-core edge and within the
semi-convective zone in the final converged models.
In addition, the mixing length in our evolution code was calibrated by
requiring that a solar model reproduce the solar luminosity and radius
as well as the solar $Z/X$ ratio at an age of 4.6 Gyr.

\section{Analysis}

\subsection{Empirical Comparisons}

For our analysis, we will use the HB distribution of NGC~2808 as an
empirical template for inspecting the HB distributions of the other
clusters in the survey.  This might seem like an unusual choice,
because NGC~2808 is not at all representative of the Milky Way
globular cluster system, being one of the most massive ($M_V =
-9.4$~mag; Harris 1996), with distinct MS and RGB sequences (D'Antona
et al.\ 2005; Piotto et al.\ 2007; Milone et al.\ 2015b) 
and prominent HB gaps (Bedin et
al.\ 2000).  However, the HB of NGC~2808 is well-populated across the
full range of temperature, with stars in the red clump, BHB, EHB, and
blue hook (Figure~2).  If any of its HB features are the result of universal
atmospheric phenomena, as opposed to a product of the multiple
populations driving the MS and RGB splitting, then these features should
align with those present in the HB distributions of other clusters, as
long as the corresponding HB locations are populated.

Given the prominence of the discontinuities 
in the CCP of Figure~2,
it is useful to compare the HB distribution of
each cluster to that of NGC~2808 in this plane.  Distance is not a factor in a
CCP, but extinction is; we need to align the HB
distributions with a fiducial that is independent of the HB
discontinuities we are investigating.  
We use the $C_{\rm peak}$ for this purpose,
which is well-populated in nearly all of our clusters.
The only exceptions are 9 clusters with
HB stars lying entirely in the red clump; for completeness, we will
include them here, and align them to NGC~2808 at the red clump, but
their lack of stars hotter than the RR~Lyrae instability strip
makes them irrelevant
to the investigation of HB discontinuities.

A possible concern when making these empirical comparisons 
to NGC~2808 is that
simple shifts within the CCP are implicitly making the
assumption that the extinction vector is independent of spectral
energy distribution (SED).  Because the photometry here covers a broad
wavelength range from the near-UV to the $B$ band, and because the HB
stars in question span a broad temperature range of
$\sim$5,000--40,000~K, this assumption is not entirely correct, but in
fact the approximation is sufficient for the purpose of making
empirical comparisons between the clusters.  This is true even though
our survey includes halo clusters with almost no foreground reddening
and bulge clusters with significant reddening. The difference in
foreground reddening for any cluster of our survey, when compared to
that of NGC~2808, can be as large as 0.5~mag in $E(B-V)$.  To
demonstrate the validity of this approximation, we show in Figure~3 stellar
structure models for the zero-age HB (ZAHB) in NGC~2808 (Brown et
al.\ 2010), which assumed [Fe/H]=$-1.36$ (Walker 1999) 
and no He enhancement\footnote{Enhancement is relative
to stars born on the MS with a primordial He abundance of $Y=0.23$.}
(i.e., $\Delta Y$~=~0), transferred to the
CCP using the LTE synthetic spectra of Castelli \&
Kurucz (2003) and the WFC3 bandpass throughputs.  In one case ({\it
  solid blue curve}), unreddened synthetic spectra were used to transfer the 
structure models to the observable plane,
and then shifted to align at the
$C_{\rm peak}$. In the other case ({\it dashed
  red curve}), the structure models were transferred to the observable
plane using spectra that were reddened with the Fitzpatrick (1999)
extinction curve, assuming $E(B-V)$~=~0.5~mag, and then shifted to align
at the $C_{\rm peak}$.  
The deviation of the observations below the theoretical ZAHB curves
between the G-jump and M-jump, due to radiative levitation,
will be explored extensively below.
The two curves agree with each other
perfectly in the middle of the HB range (by definition), but separate
as one looks to the red and blue extremes of the HB.  However, the
distinctions between the models are small: 
at the location of the G-jump, using
either the red curve or the blue curve to estimate the effective temperature
of the G-jump would only change the estimate by 100~K
(0.02~mag change in $m_{\rm F275W} - m_{\rm F438W}$),
while at the location of the M-jump, the
temperature estimates would only differ by 500~K
(0.04~mag change in $m_{\rm F275W} - m_{\rm F438W}$).
Thus, even when comparing HB distributions of clusters with significantly
distinct foreground reddening, the agreement in the colors of these
features implies agreement in their temperatures at the level of
$\sim$500~K or less.  In cases where the colors of these
discontinuities do not agree between clusters, their temperatures can
be quantified with a theoretical ZAHB distribution tailored to match
the cluster in question, including an SED-dependent foreground extinction.

%fig3
\begin{figure}[t]
\includegraphics[width=3.4in]{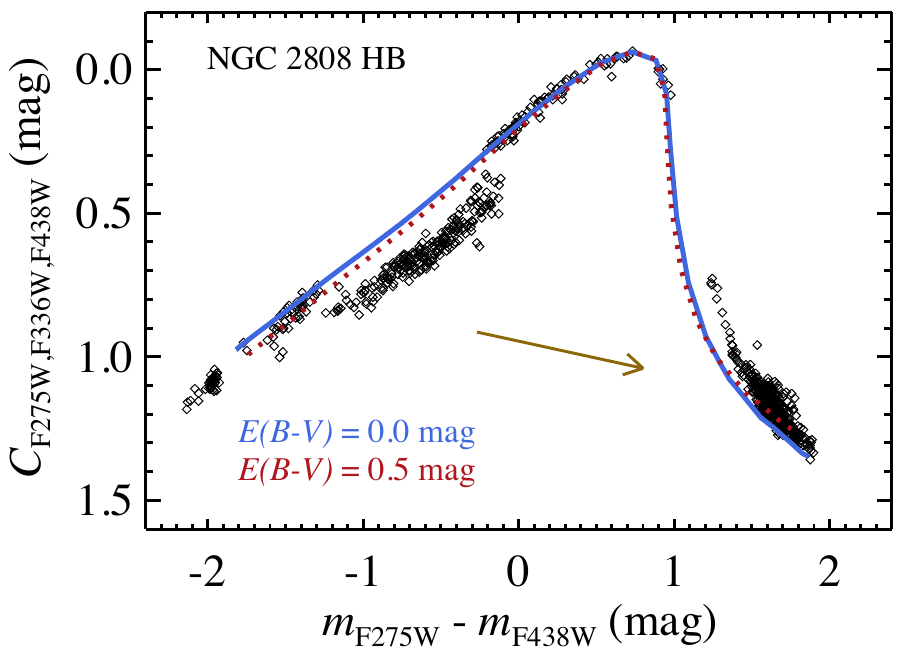}
\caption{The observed HB distribution of NGC~2808 ({\it black points}) in the
CCP that highlights its HB discontinuities.  
The extinction vector for a 12,000~K BHB star is shown ({\it brown arrow}).
Strictly speaking,
the extinction vector in this plane is SED dependent. However, for
the purposes of empirical comparisons to the other clusters in our sample, 
one can assume an SED-independent
extinction vector and simply shift the HB distribution of any cluster 
to align with that of NGC~2808, using the
$C_{\rm peak}$ as a fiducial. We demonstrate the accuracy of this approach 
by shifting a theoretical ZAHB distribution 
to align with the observations at the
$C_{\rm peak}$, using two different extinction assumptions.  
In one case, no foreground reddening was applied before the 
alignment ({\it solid blue curve}),
and in the other, significant SED-dependent reddening 
was applied before the alignment
({\it dotted red curve}), but the distinction between the two
curves is small.  The deviation of the observed BHB below the ZAHB curves
at $-1.2 < (m_{\rm F275W} - m_{\rm F438W}) < -0.2$~mag is due to radiative 
levitation blueward of the G-jump.  The stars falling
to the blue of the ZAHB curves, at $(m_{\rm F275W} - m_{\rm F438W}) < -1.8$~mag,  
are blue-hook stars, with effective temperatures
hotter than the canonical end of the EHB.}
\end{figure}

%fig4
\begin{figure*}[t]
\begin{center}
\includegraphics[width=6.0in]{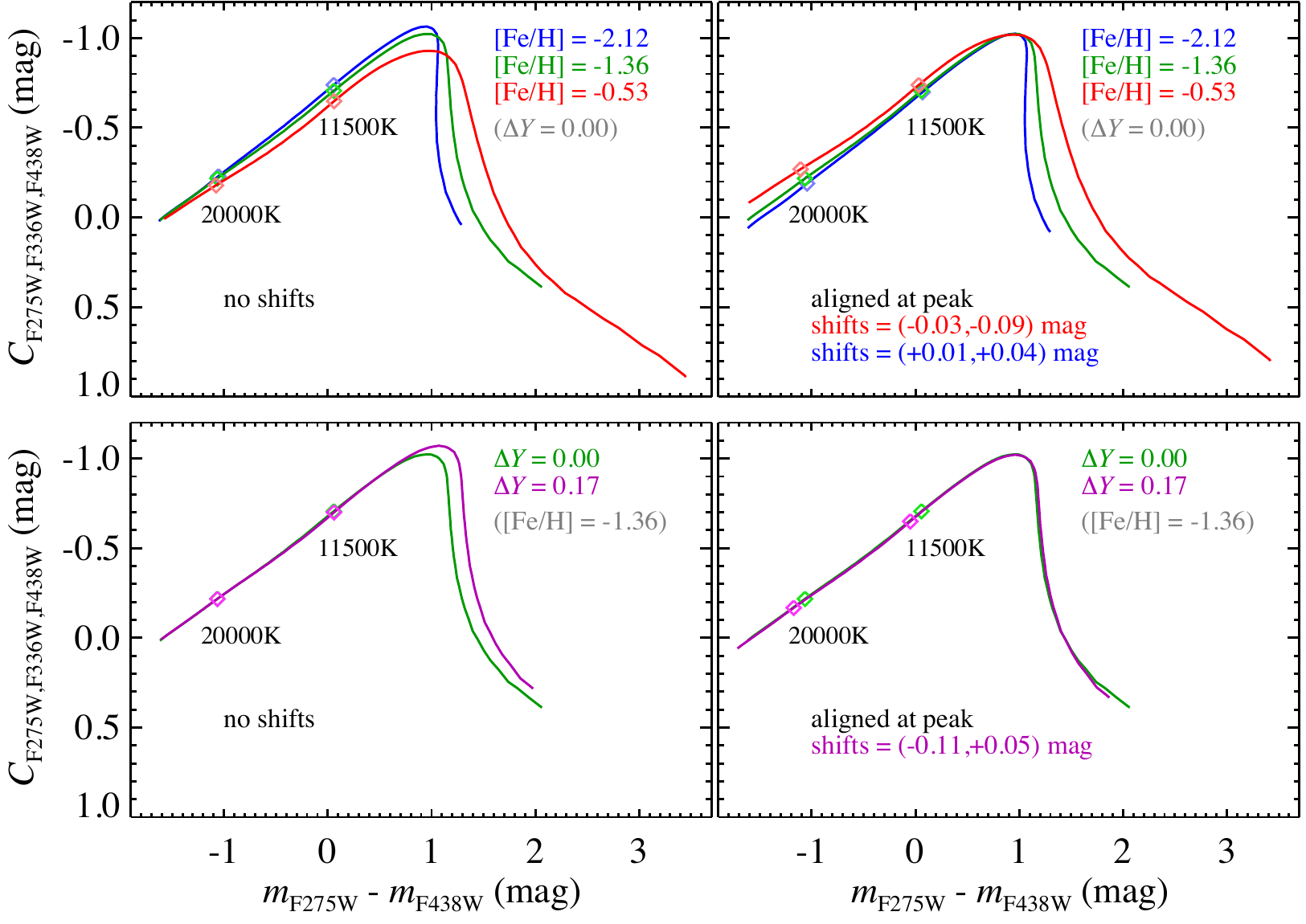}
\end{center}
\caption{
  Theoretical ZAHB models ({\it curves}) before ({\it left panels}) and
  after ({\it right panels}) they are aligned at the $C_{\rm peak}$, for
  variations in metallicity ({\it top panels}) and He abundance ({\it bottom
  panels}). The $C_{\rm peak}$ reference point on the BHB falls at 
  8,600$\pm 100$~K, and this holds true
  over the range of $-0.53 \leq $~[Fe/H]~$-2.12$ and over the range 
  of $0 \leq \Delta Y \leq 0.17$.  Although NGC~2808 is an unusually
  massive cluster, its abundance profile and well-populated HB make 
  it a useful template population;
  the BHB stars of NGC~2808 fall in the middle of these abundance ranges,
  with intermediate metallicity ([Fe/H]~=~$-1.36$; Walker 1999),
  and moderate He enhancement ($\Delta Y \sim 0.09$; Marino et al.\ 2014).
  Using $C_{\rm peak}$ to align 
  the HB distribution of any cluster in our sample to that of NGC~2808
  should give agreement at the G-jump and M-jump, at the level of 0.06~mag
  (or better) in $m_{\rm F275W} - m_{F438W}$ color,
  unless the temperatures of these features are not universal.
}
\end{figure*}

Another possible concern when comparing observed HB distributions
to that of NGC~2808 is the effect of abundance distinctions between
clusters.  Fortunately, ZAHB models demonstrate 
that the intrinsic $m_{\rm F275W} - m_{\rm F438W}$ color
is not very sensitive to $Y$ or [Fe/H] at the
effective temperatures of the $C_{\rm peak}$ ($\sim$8,600~K),
the G-jump ($\sim$11,500~K), or the M-jump ($\sim$20,000~K),
as demonstrated in Figure~4.  In the top panels, we show
the effects of [Fe/H]; in the bottom panels, we show the effects of 
$Y$.  In the left panels, we show the ZAHB distributions at their relative
positions in the CCP, before any alignments are made; in the right panels,
we show the ZAHB distributions after they have been aligned at the 
$C_{\rm peak}$ reference point.  Because NGC~2808 is of intermediate metallicity 
([Fe/H]~=~$-1.36$; Walker 1999), 
using the $C_{\rm peak}$ to align clusters of either 
high or low metallicity incurs misalignments in
$m_{\rm F275W} - m_{\rm F438W}$ color of less than 0.05~mag at the G-jump 
and M-jump.  
Because the BHB stars near the $C_{\rm peak}$ in NGC~2808
are only moderately enhanced in He ($\Delta Y \sim 0.09$; Marino et al.\ 2014),
using the $C_{\rm peak}$ to align clusters with little He enhancement
or strong He enhancement
incurs misalignments in
$m_{\rm F275W} - m_{\rm F438W}$ color of less than 0.06~mag at the G-jump 
and M-jump, which corresponds to a $T_{\rm eff}$ difference of 240~K
at the G-jump and 740~K at the M~jump. If the systematic errors from
[Fe/H] and $Y$ went in the same direction, the misalignment could be
as large as 0.1~mag.

In Figure~5, we compare the HB distribution of NGC~2808 ({\it grey
  points}) to that of each of the other 52 clusters in our sample
({\it blue points}).  Each of the HB distributions has been aligned
to that of NGC~2808 at the $C_{\rm peak}$, accounting for distinctions
in composition and extinction.
To aid these comparisons, the 52 clusters have
been sorted 
into five arbitrary categories (explained below) 
with increasingly blue HB morphology.  Within each category, we also
sort the clusters
by the mean $(m_{\rm F275W} - m_{\rm F438W})$ color of the 10
bluest HB stars in each cluster, after the cluster HB has been aligned
to that of NGC~2808 (otherwise it would depend upon both HB morphology
and extinction).
The choice of sorting metric is arbitrary, but
our chosen metric is more useful than other obvious choices, such as
the color of the single bluest star (which is hampered by outliers) or
the mean color of the entire HB (which is affected by strongly bimodal
HB distributions). 
Of the 52 clusters compared to NGC~2808, 17 do not
host sufficiently blue HB stars to characterize the three HB discontinuities
blueward of the RR~Lyrae instability strip.  Of the remaining 36 clusters,
all but two (NGC~6388 and NGC~6441) exhibit excellent agreement with the
discontinuities observed in NGC~2808, although the varying HB morphologies
of these clusters do not always lend themselves to exploring each feature
in detail.  Furthermore, the red clump in these clusters can vary significantly
from that in NGC~2808, as will be discussed later.

Our data do not provide sufficient time sampling to identify and
characterize RR~Lyrae stars.  However, candidate RR~Lyrae stars can be
flagged by large exposure-to-exposure photometric variations.
In Figure~5, we have excluded candidate RR~Lyrae stars by omitting 
stars that exhibit photometric variations that significantly
exceed those of other
stars at the same magnitude (at the level of 5$\sigma$ or greater).
A representative example of the 
photometric uncertainty is indicated
by an error bar in the upper right-hand corner of each panel.
Because the RR~Lyrae instability strip is redder than
the discontinuities we explore in this paper, the exclusions do not
affect our analysis, but avoid cluttering the CCP with variable stars
sampled at random phases.

%fig5a
\begin{figure*}[p]
\begin{center}
\includegraphics[width=6.5in]{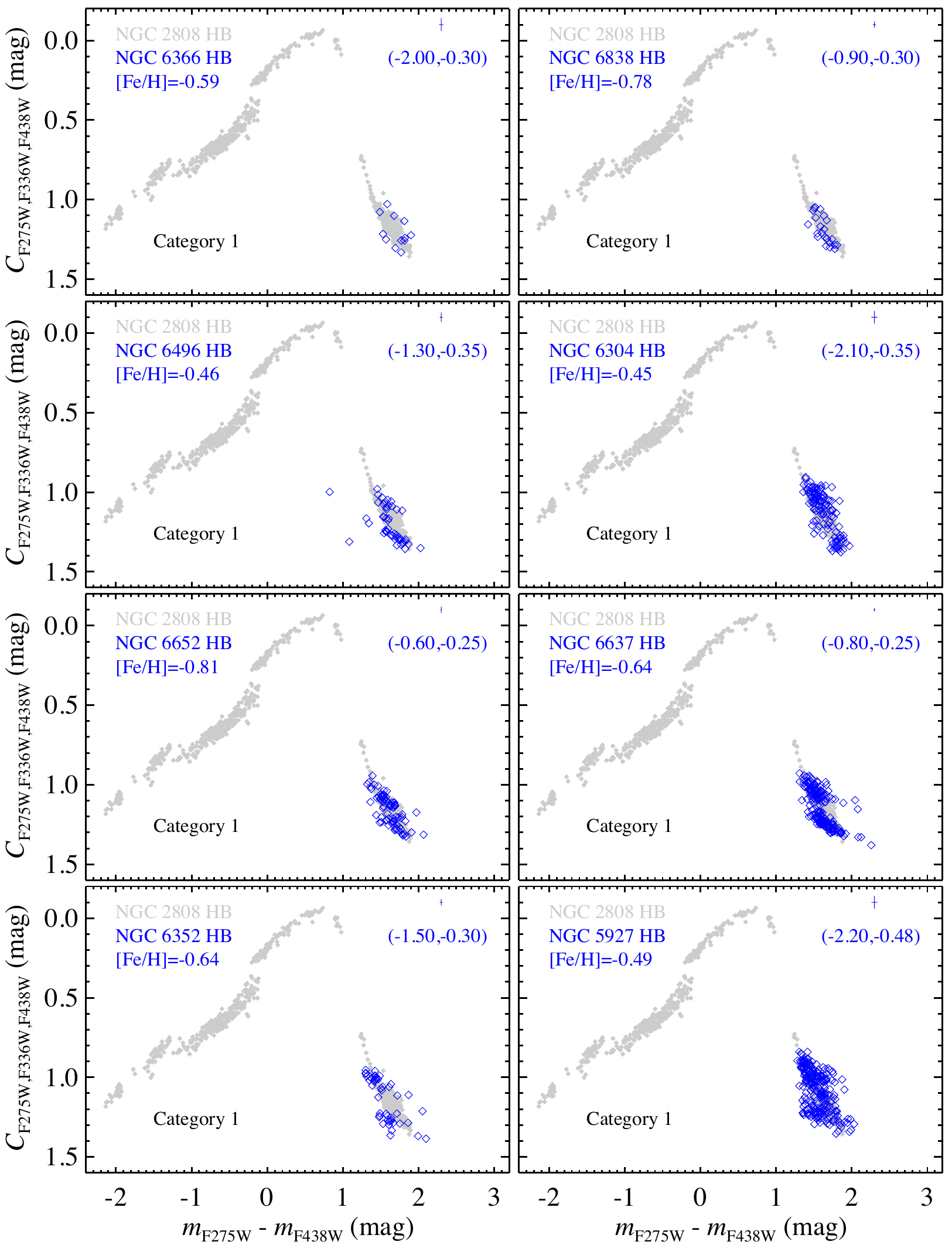}
\end{center}
\caption{The HB distributions of each globular cluster in our sample
({\it blue points}) compared to that of NGC~2808 ({\it grey points}).
The HB distribution of each cluster
has been aligned with that of NGC~2808 at the
$C_{\rm peak}$, using the 
shifts indicated in parentheses.  
To aid comparisons, the clusters
are sorted into arbitrary 
categories of increasingly blue HB morphology (see text).
The metallicity of each cluster is indicated (Harris 1996; Brown et al.\ 2010).
Representative photometric errors are shown ({\it error bar} in each panel).
Candidate RR~Lyrae stars have been excluded.
}
\end{figure*}

%fig5b
\addtocounter{figure}{-1}
\begin{figure*}[p]
\begin{center}
\includegraphics[width=6.5in]{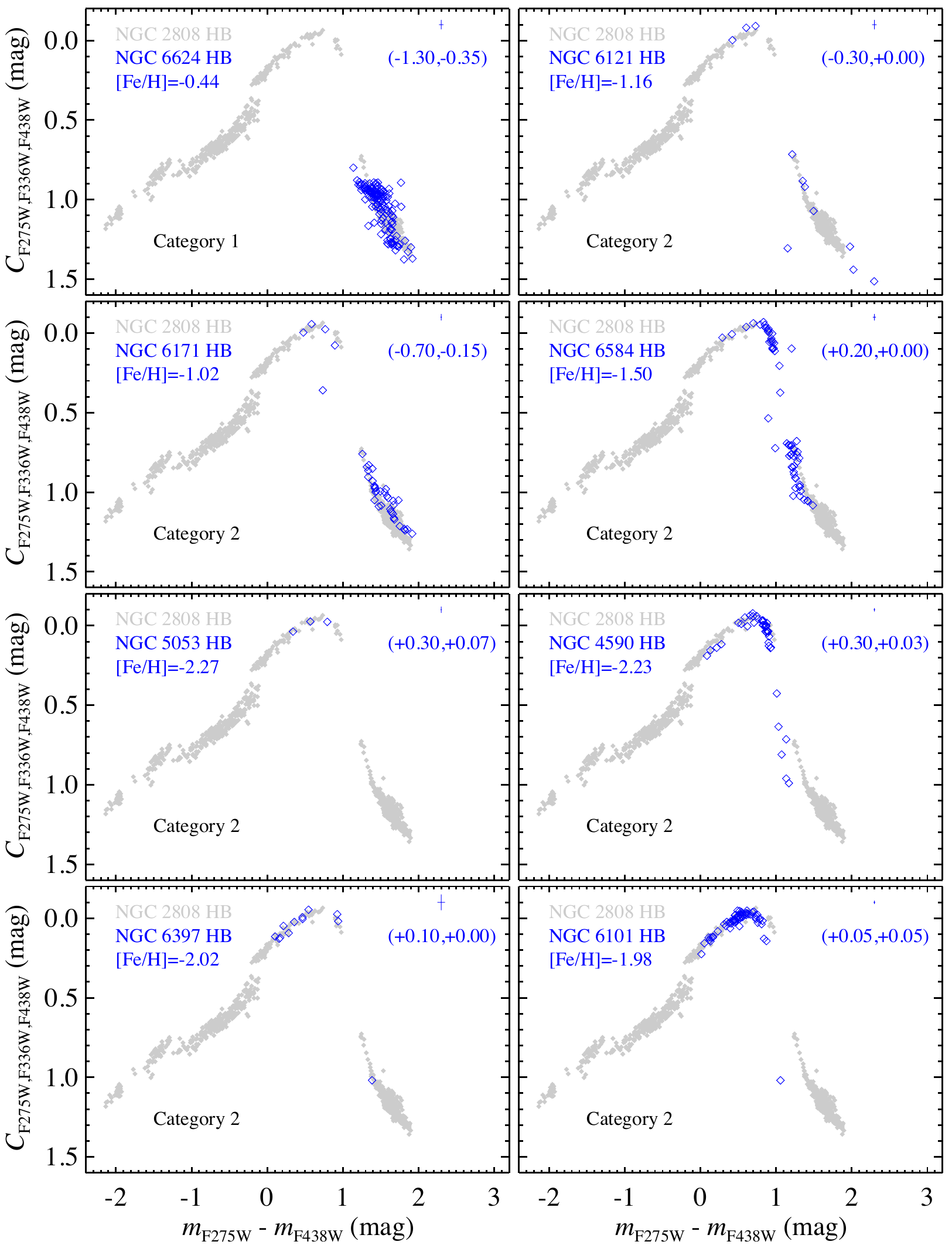}
\end{center}
\caption{{\it Continued} \\ \\}
\end{figure*}

%fig5c
\addtocounter{figure}{-1}
\begin{figure*}[p]
\begin{center}
\includegraphics[width=6.5in]{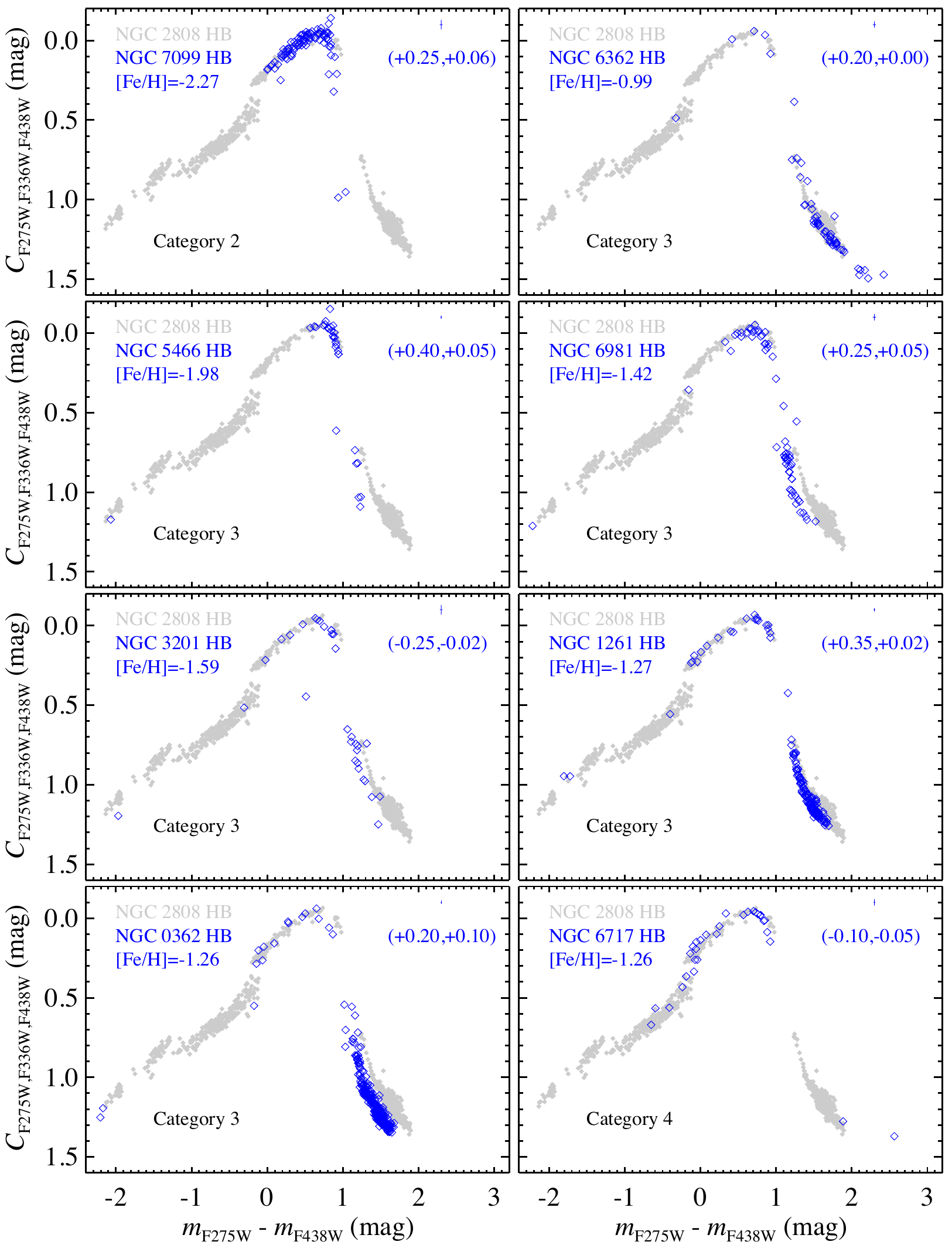}
\end{center}
\caption{{\it Continued} \\ \\}
\end{figure*}

%fig5d
\addtocounter{figure}{-1}
\begin{figure*}[p]
\begin{center}
\includegraphics[width=6.5in]{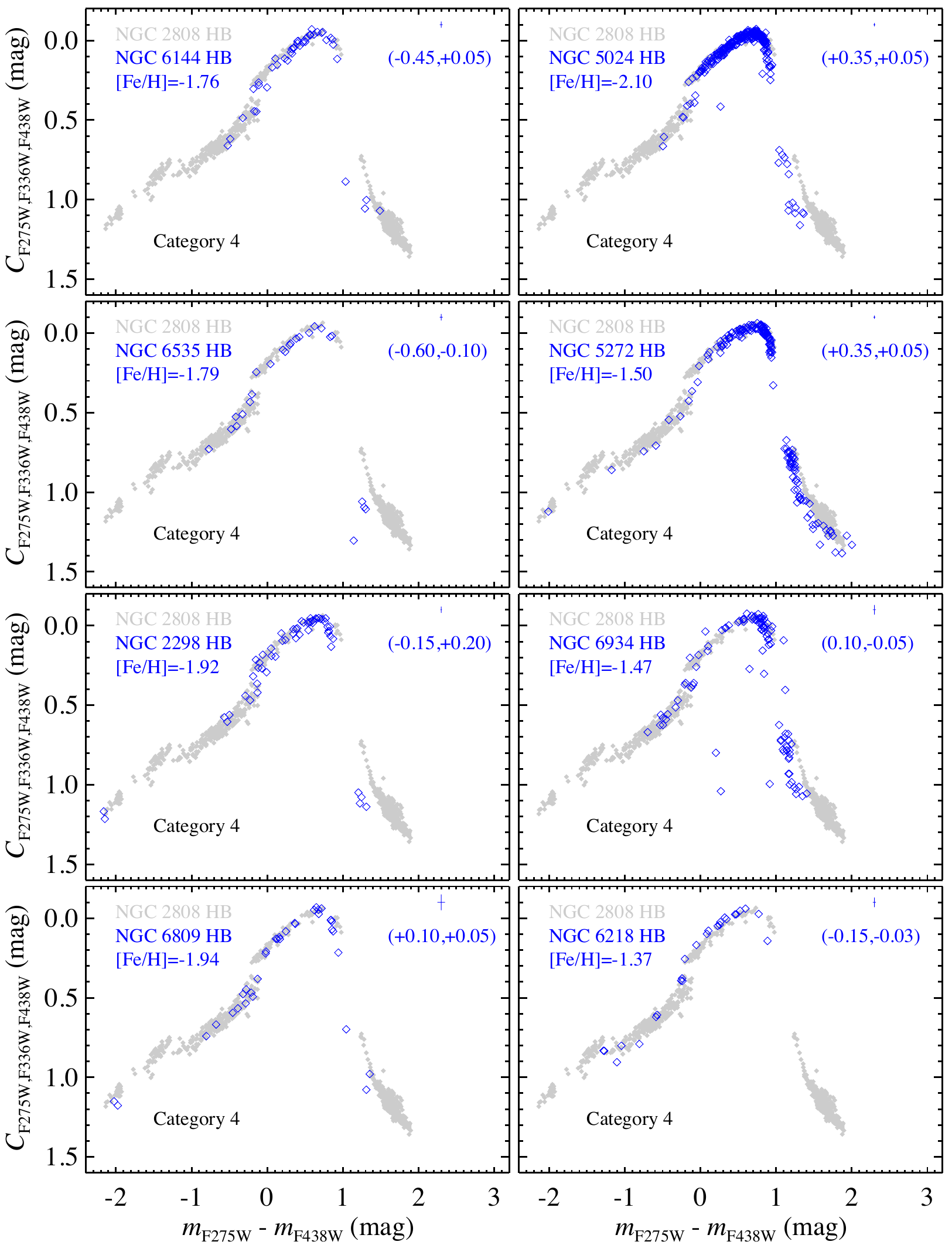}
\end{center}
\caption{{\it Continued} \\ \\}
\end{figure*}

%fig5e
\addtocounter{figure}{-1}
\begin{figure*}[p]
\begin{center}
\includegraphics[width=6.5in]{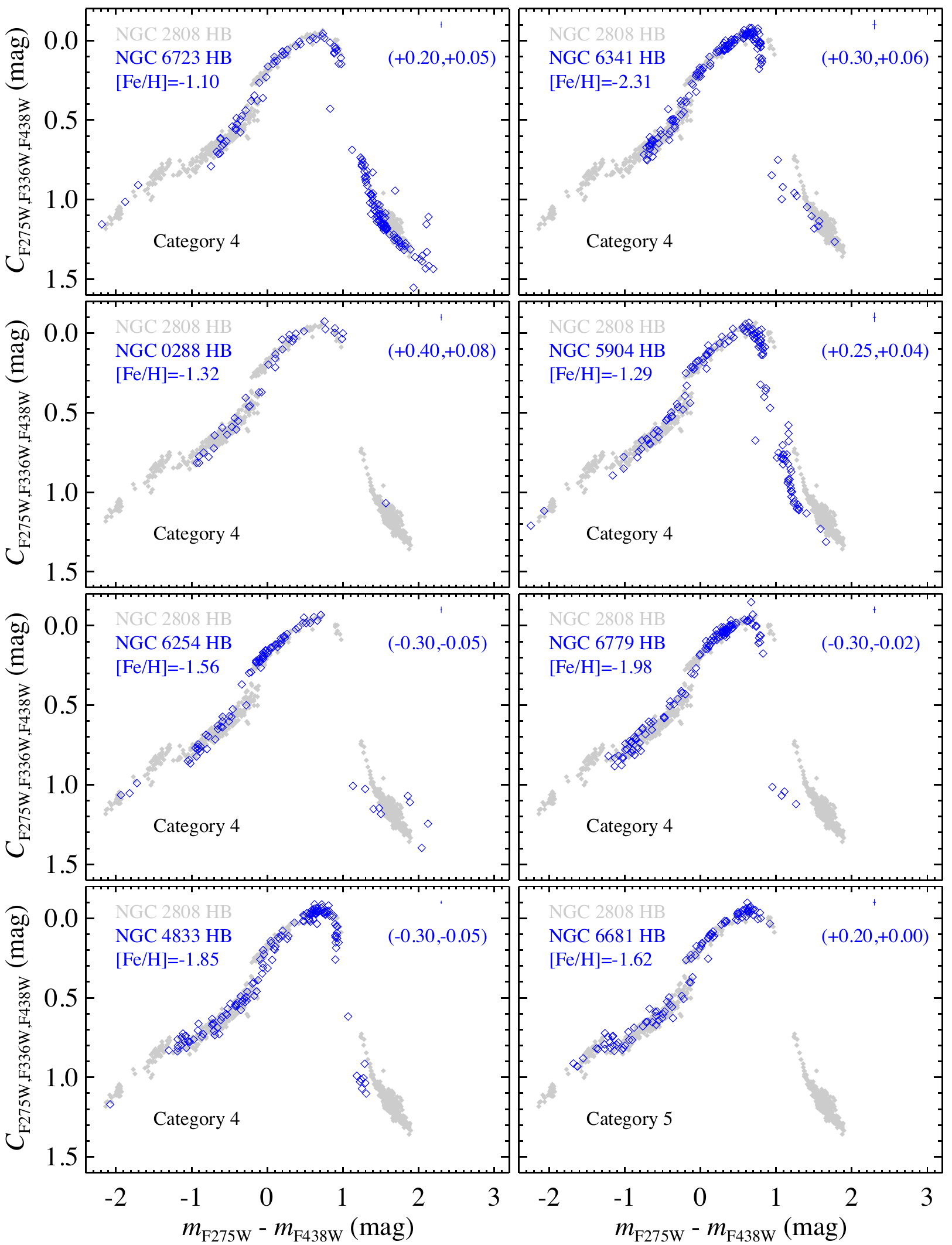}
\end{center}
\caption{{\it Continued} \\ \\}
\end{figure*}

%fig5f
\addtocounter{figure}{-1}
\begin{figure*}[p]
\begin{center}
\includegraphics[width=6.5in]{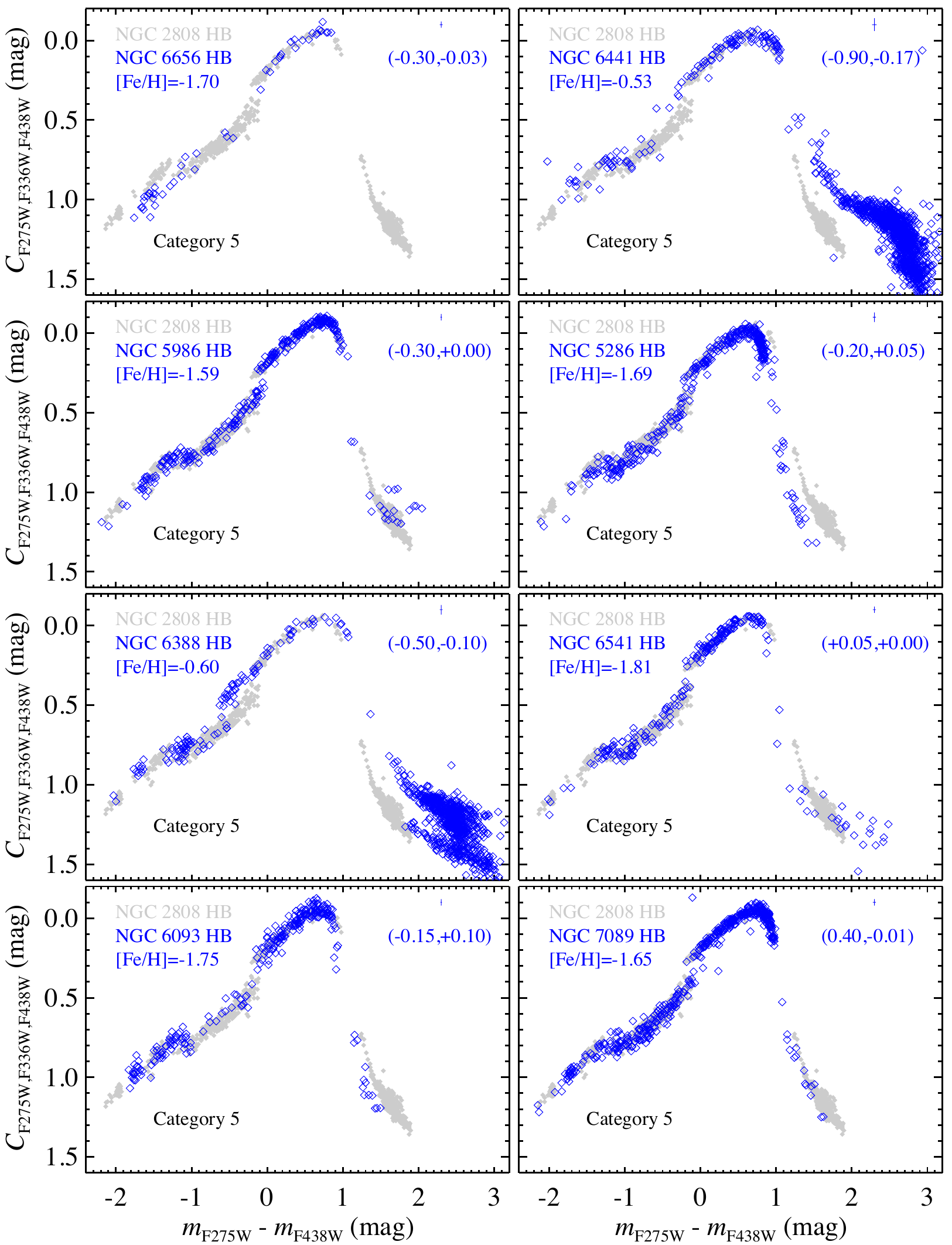}
\end{center}
\caption{{\it Continued} \\ \\}
\end{figure*}

%fig5g
\addtocounter{figure}{-1}
\begin{figure*}[t]
\begin{center}
\includegraphics[width=6.5in]{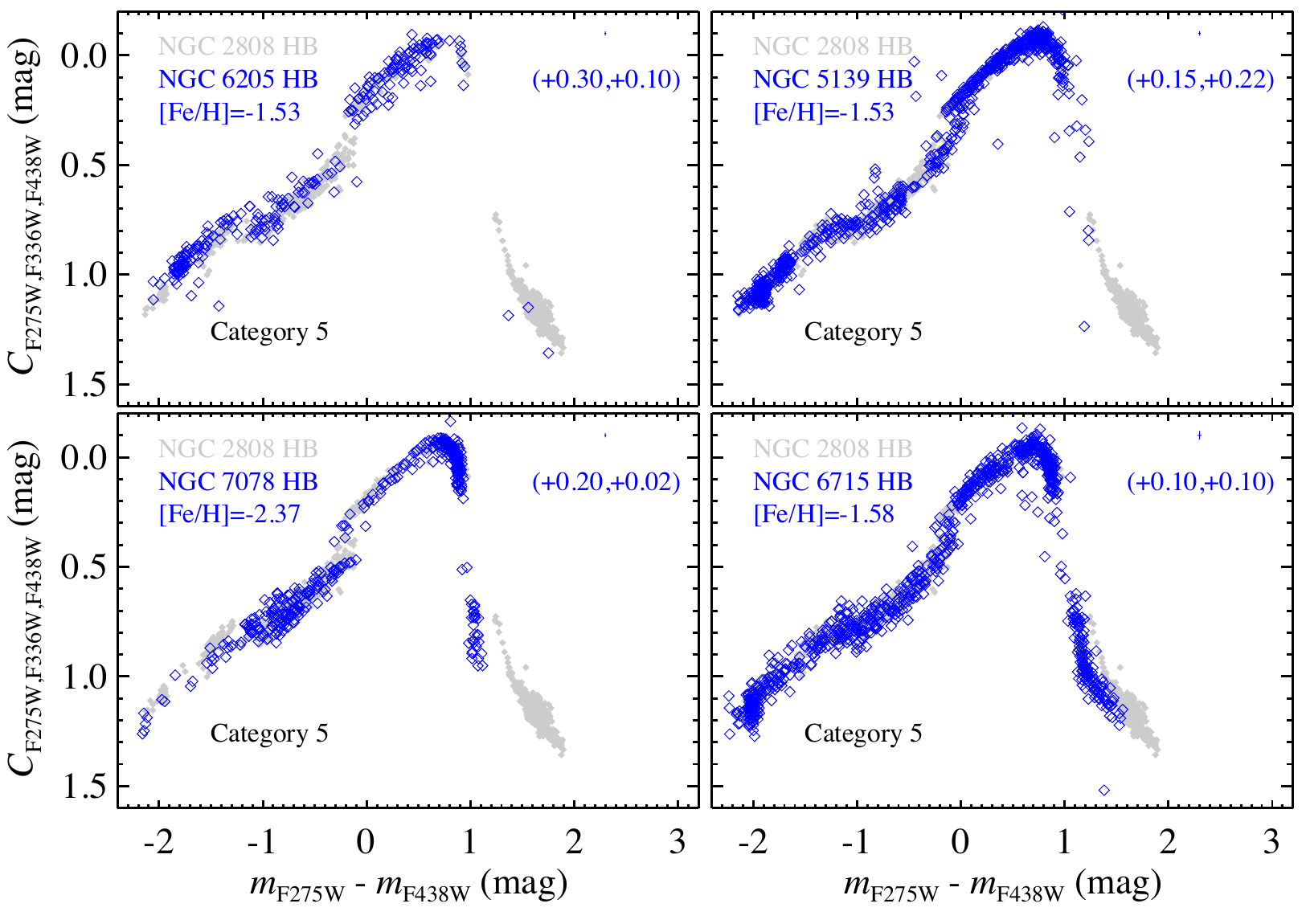}
\end{center}
\caption{{\it Continued}}
\end{figure*}

\noindent
{\it Category 1:} There are 9 clusters dominated by red clump stars
(NGC~6366, NGC~6838, NGC~6496, NGC~6304, NGC~6652, NGC~6637, NGC~6352, 
NGC~5927, and NGC~6624).  These
clusters provide no information on any of the HB discontinuities.

\noindent
{\it Category 2:} The HB distributions in 8 additional clusters 
(NGC~6121, NGC~6171, NGC~6584, NGC~5053, NGC~4590, NGC~6397, NGC~6101, 
and NGC~7099) extend
through the RR~Lyrae instability strip, but not far enough to populate
the vicinities of the Grundahl and Momany jumps.  Again, these clusters
provide no information on the HB discontinuities.

\noindent
{\it Category 3:} In 6 cases (NGC~6362, NGC~5466, NGC~6981, NGC~3201, 
NGC~1261, and NGC~362), there are less than four HB stars observed
blueward of the G-jump, but those stars do in fact trace the
deviations observed in the HB distribution of NGC~2808. In 4 of these
clusters (NGC~362, NGC~5466, NGC~6981, and NGC~3201) there are a few
blue-hook stars.  However, given the scarcity of BHB and EHB stars, 
these clusters do not provide interesting constraints on the 
colors of the HB discontinuities.

\noindent
{\it Category 4:} There are 16 clusters where the region between the
Grundahl and Momany jumps is significantly populated, with few stars
hotter than this region.  All of the BHB stars trace the
HB deviations observed in NGC~2808, and do so with enough stars
to demonstrate consistency in the colors of the Grundahl and 
Momany jumps.  In 9 cases (NGC~6717, NGC~6144, NGC~5024, NGC~6535,
NGC~6934, NGC~6218, NGC~6341, NGC~288, and NGC~6779), 
there are no stars hotter than the M-jump.  In 7 more
clusters (NGC~5272, NGC~2298, NGC~6809, NGC~6723, NGC~5904, NGC~6254, 
and NGC~4833), those few stars hotter than the M-jump include blue-hook
stars.  Some of these clusters (e.g., NCC~5904, NGC~6779, NGC~6254,
and NGC~4833), exhibit a sharp decline in the density of HB stars 
at the M-jump, but this is likely a coincidence, given that
other clusters in this category (e.g., NGC~6341 and NGC~288)
exhibit sharp declines unassociated with any particular HB discontinuity.

\noindent
{\it Category 5:} There are 14 clusters (NGC~6681, NGC~6656, NGC~6441,
NGC~5986, NGC~5286, NGC~6388, NGC~6541, NGC~6093, NGC~7089, NGC~6205,
NGC~5139, NGC~7078, NGC~6715, and NGC~2808 itself) 
that have significant numbers of stars
straddling both the Grundahl and Momany jumps.  Most of these also
host blue-hook stars, with the exception of NGC~6681, NGC~6656, 
NGC~6441\footnote{NGC~6441 hosts a
population of subluminous HB stars that are much redder than the
blue-hook stars found in other clusters, so their status is unclear
(Brown et al.\ 2010).}, and
NGC~6093, which truncate in the gap between the blue-hook stars and
the normal EHB population.  NGC~6656 (M22) looks somewhat unusual blueward
of the M-jump, where the $C_{\rm F275W, F336W, F438W}$ index does not completely
overlap with that of NGC~2808.  In most of these 14 clusters, the HB
discontinuities have the same $m_{F275W} - m_{F438W}$ colors as those
in NGC~2808; the exceptions are NGC~6388 and NGC~6441, which each
exhibit a G-jump that is significantly bluer than normal.  
The red clump distributions of NGC~6388 and NGC~6441 also
appear very distinct from those in the other clusters, with NGC~6388
having a bifurcated structure (see also Bellini et al.\ 2013b).  \\ \\

\subsection{Clusters with Blue-Hook Stars}

Our sample 
includes 21 clusters with blue-hook stars, extending
to 23 the tally of globular clusters
known to host blue-hook stars. 
In Table 1, we list these clusters, along with the 
total number of HB stars in each sample ($n_{\rm HB}$), and the 
total number of blue-hook stars ($n_{\rm BH}$) that are blue enough to be
unambiguously classified.  Specifically, we classify them
as blue-hook stars if they fall
within $-2.4 \le m_{\rm F275W} - m_{\rm F438W} \le -1.9$~mag and
$1.0 \le C_{\rm F275W,F336W,F438W} \le 1.3$~mag in the CCP after alignment
with NGC~2808.  Note that there 
are clusters in our sample where a few additional
stars fall immediately to the red of this selection region, which might
also be blue-hook stars, but their classification would not be secure
(NGC~5139, NGC~5286, NGC~5986, NGC~6205, NGC~6254, NGC~6541, 
NGC~6715, NGC~6723, and NGC~7089).

\begin{table}[t]
\begin{center}
\caption{Survey Clusters with Blue-Hook Stars}
\begin{tabular}{rrr|rrr|rrr}
\tableline
NGC &  $n_{\rm HB}$ & $n_{\rm BH}$ & NGC &  $n_{\rm HB}$ & $n_{\rm BH}$ & NGC &  $n_{\rm HB}$ & $n_{\rm BH}$ \\
\tableline
 362 & 303 &   2 &  5286 &  386 &   2 &  6541 &  228 &   3 \\
2298 &  56 &   2 &  5466 &   30 &   1 &	 6715 &  916 & 120 \\
2808 & 757 &  52 &  5904 &  132 &   2 &	 6723 &  150 &   1 \\
3201 &  31 &   1 &  5986 &  328 &   3 &	 6809 &   35 &   2 \\
4833 & 138 &   1 &  6205 &  244 &   4 &	 6981 &   64 &   1 \\
5139 & 825 & 156 &  6254 &   86 &   1 &	 7078 &  390 &   7 \\
5272 & 179 &   1 &  6388 &  953 &   2 &	 7089 &  515 &   3 \\
\tableline
\end{tabular}
\end{center}
\end{table}

To date, UV photometry had been used to
confirm blue-hook stars in six globular clusters: NGC~5139 ($\omega$
Cen), NGC~2808, NGC~6715 (M54), NGC~2419, NGC~6388, and NGC~6273
(D'Cruz et al.\ 2000; Brown et al.\ 2001; Dalessandro et al.\ 2008;
Dieball et al.\ 2009; Brown et al.\ 2010).  Although our sample does
not include NGC~2419 and NGC~6273, 
it does include the others, and the CCP
of Figure~5 confirms the presence of blue-hook stars in each
cluster.  Dieball et al.\ (2009, 2010) provided
four additional clusters that
may host a small number of blue-hook stars each: NGC~6093 (M80),
NGC~6681 (M70), NGC~7078 (M15), and NGC~6441.
Of these clusters, three (NGC~6093,
NGC~6441, and NGC~6681) do not however appear to host blue-hook stars in our
sample, but NGC~7078 clearly does.  We also find blue-hook
stars in 16 additional clusters: 
NGC~362, NGC~2298, NGC~3201, NGC~4833, NGC~5272 (M3), 
NGC~5286, NGC~5466, NGC~5904 (M5), NGC~5986, NGC~6205 (M13),  
NGC~6254 (M10), NGC~6541, NGC~6723, NGC~6809 (M55), NGC~6981 (M72), 
and NGC~7089 (M2). In 7 of these clusters 
(NGC~3201, NGC~4833, NGC~5272, NGC~5466, NGC~6254, NGC~6723, and NGC~6981),
there is only a single
star falling unambiguously in the blue-hook region.
The classification seems secure, given 
the placement in both the CMD and CCP,
and the lack of other stars in the vicinity that would otherwise suggest
significant contamination.  In the 23 clusters hosting blue-hook stars,
these stars comprise up to 20\% of the HB population,
although the percentage
is highest in three of the most massive clusters 
(NGC~2808, NGC~5139, and NGC~6715).
Our sample of 53 clusters spans $-9.98 \le M_V \le -4.75$~mag in total
luminosity (Harris 1996), with a median of $-7.48$~mag; 17 of the 21 clusters in
our sample hosting blue-hook stars are in the brightest half of the
sample, reinforcing the idea that blue-hook stars tend to form in the
most massive clusters (see Brown et al.\ 2010 for a full discussion).
For some of these clusters, the statistics are poor, with only one or two
blue-hook stars, but we stress that the presence of blue-hook stars
is not simply a matter of sampling enough stellar mass to find a relatively
rare evolutionary phase.  For example,
the mean metallicities of NGC~5139 and NGC~5986 match within 0.1~dex, and
the counts of their HB stars in our catalogs match within 3\%, but NGC~5139,
being the more massive cluster,
has $>$20 times as many blue-hook stars as NGC~5986;
this is another manifestation of parameters beyond metallicity 
affecting the HB morphology (i.e., the second-parameter problem).

\subsection{The Grundahl and Momany Jumps}

\subsubsection{NGC~6388 and NGC~6441}

In Figure~5, only two globular clusters exhibit significant
discrepancies with the discontinuities of NGC~2808: NGC~6388 and NGC~6441.
Specifically, in both of these clusters, the G-jump is
$\sim$0.4~mag bluer than that
observed elsewhere.  In NGC~6388, the region between the G-jump and
M-jump is well populated, such that the shift is obvious.  In
NGC~6441, the shift is less statistically significant, with only 6
stars falling blueward of the point where the G-jump occurs in
NGC~2808, and only 2 of these stars at significantly bluer colors.
That said, all 6 of these stars are well aligned with the $C_{\rm
  F275W,F336W,F438W}$ trend extending from the cooler HB stars in NGC~6441, and
over the $m_{F275W} - m_{F438W}$ color range of these 6 stars, there are
no stars exhibiting a deviation from this trend (i.e., these 6 stars do
not trace the locus of NGC~2808).  In both NGC~6388 and NGC~6441, the
shift in this feature implies that the onset for radiative levitation
occurs at hotter effective temperature than the onset in other clusters,
even after one accounts for the composition effects demonstrated in Figure~4.

%fig6
\begin{figure*}[t]
\begin{center}
\includegraphics[width=6.5in]{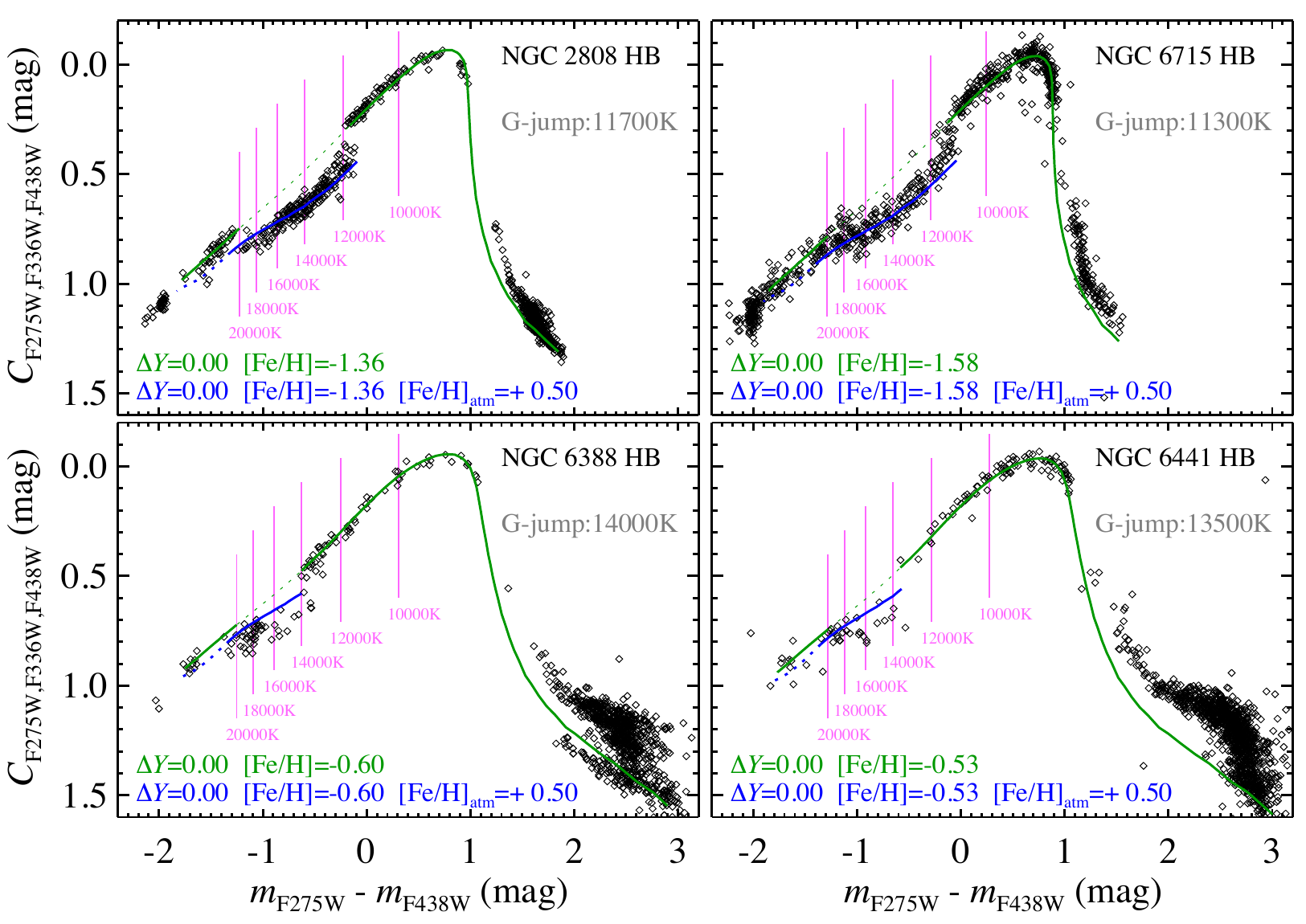}
\end{center}
\caption{The HB distributions for 4 massive clusters in our sample
  ({\it black points}), compared to theoretical ZAHB distributions matching the 
  cluster metallicity and assuming $\Delta Y$~=~0 
  ({\it blue and green curves}).  
  An effective temperature scale is shown to guide the 
  eye ({\it pink lines}; labeled).
  If the ZAHB stellar structure models are transferred to the CCP
  using synthetic spectra representative of the cluster
  composition ({\it green curves}, 
  shown as a {\it dotted line} between the G-jump and M-jump),
  the observed deviation between the
  Grundahl and Momany jumps is obvious.  If the ZAHB stellar structure models
  are transferred to the CCP using synthetic
  spectra of enhanced metallicity ({\it blue curves},
  shown as a {\it dotted line} beyond the M-jump),
  the model tracks the deviation between the G-jump and M-jump.  Although
  the restoration of normal atmospheric abundances blueward of the M-jump
  matches the model to the data in this plane, this does not 
  reproduce the behavior in the CMD (see Figure~8).
  We can use the points where the observed HB distribution deviates
  from the standard ZAHB distribution to determine the temperature
  of the G-jump and M-jump.  For the
  intermediate-metallicity clusters NGC~2808 and NGC~6715 ({\it top panels}),
  the temperatures of those two features match those in dozens of other Galactic
  globular clusters (Figure~5).  For the 
  metal-rich clusters NGC~6388 and NGC~6441 ({\it bottom panels}),
  the G-jump is $\sim$2,000~K hotter than normal.}
\end{figure*}

%fig7
\begin{figure*}[t]
\begin{center}
\includegraphics[width=6.5in]{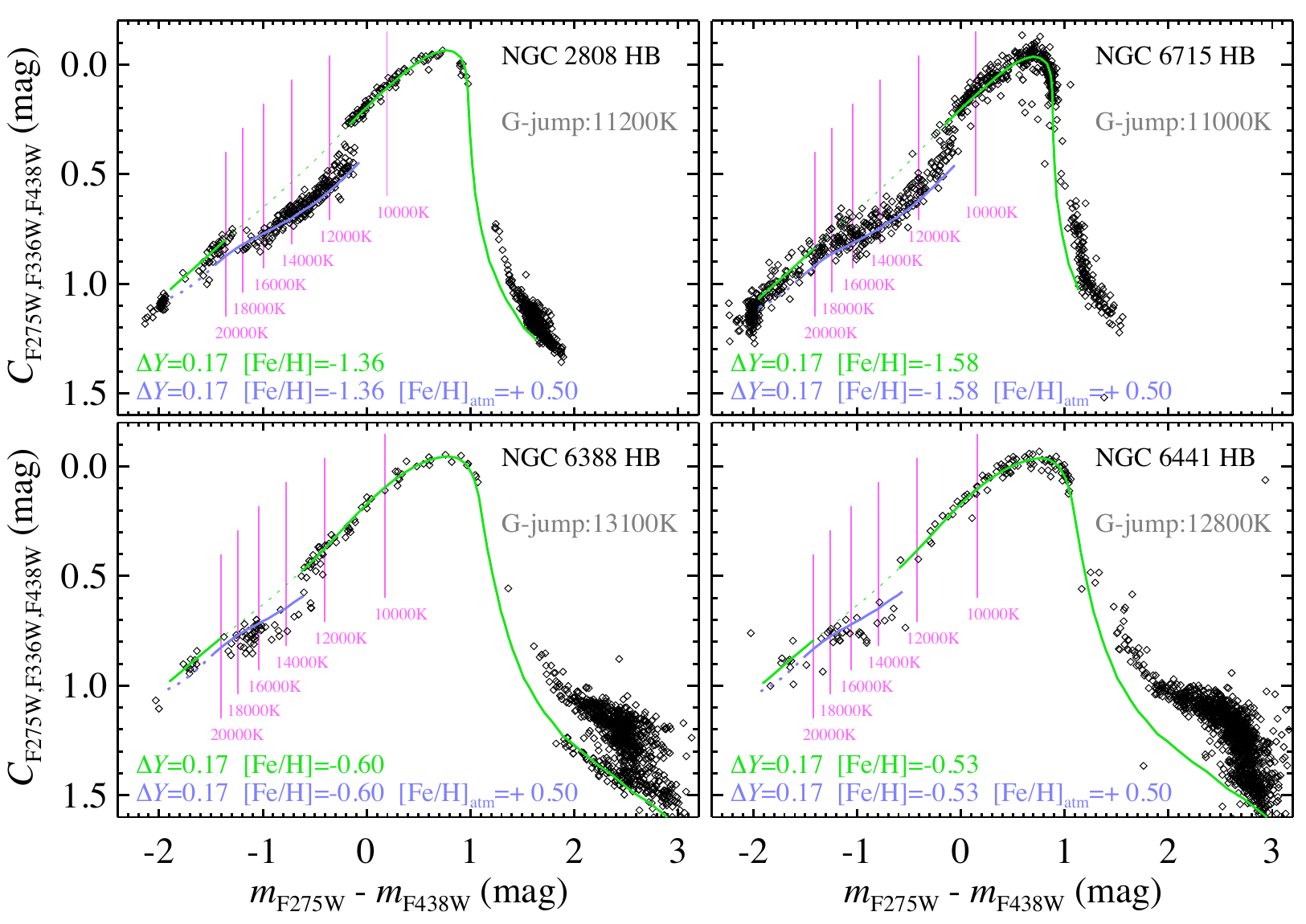}
\end{center}
\caption{
  The same as Figure~6, but with He-enhanced stellar structure models.
  Although the stellar structure models are matched to each cluster 
  in metallicity, the He abundance has a small but non-negligible
  effect on the alignment of the models to the data (see Figure~4).
  For this reason, the effective temperature of the G-jump and M-jump
  in each cluster is bracketed by the estimates here and those in Figure~6.
  The G-jump in the metal-rich clusters (NGC~6388 and NGC~6441) is 
  much hotter than that in the other clusters, even if one takes
  the coolest estimate (here) for the metal-rich clusters and the hottest
  estimate (Figure~6) for the other clusters.
}
\end{figure*}

To quantify this shift, in Figure~6 we compare the photometry of four
clusters (NGC~2808, NGC~6715, NGC~6388, and NGC~6441)  
to theoretical ZAHB distributions from Brown et al.\ (2010).
These ZAHB models assume that the HB stars evolved from MS stars with the
standard chemical composition for each cluster (i.e., no
He enhancement). The stellar structure 
models were then
transferred to the observed CCP using the LTE synthetic
spectra of Castelli \& Kurucz (2003), applying the Fitzpatrick (1999)
reddening law to provide appropriate SED-dependent extinction at each point
along the ZAHB (thus avoiding the small approximation errors
demonstrated in Figure~3), and subsequently folding the spectra through the
WFC3 bandpasses.  As in Figure~5, we aligned the theoretical distributions 
at the $C_{\rm peak}$, using a least squares fit of the
model to the data.  The observed HB distribution ({\it black points})
between the Grundahl and Momany jumps clearly deviates from the theoretical
distribution ({\it green curves}) in Figure~6; the model is shown as a
dashed line over the region where the data deviate from the model.
Over the range of the observed deviation, we also show the same stellar
structure models, but transferred to the observable plane using 
synthetic spectra with an enhanced metallicity of [Fe/H]~=~0.5
(the maximum metallicity available in the grid of synthetic spectra), simulating
the effects of radiative levitation in the atmospheres ({\it blue curves}).  
For NGC~6388 and NGC~6441, the deviation observed between the G-jump
and M-jump is more significant than that in the model.
This may be due to the fact that the stars in these
clusters were born at much higher metallicity than NGC~2808, such that
a simulation of radiative levitation would require a metallicity
exceeding [Fe/H]~=~0.5. 
Of course, our use of [Fe/H]~=~0.5 synthetic spectra is only a crude
approximation of the actual interplay between gravitational settling
and radiative levitation in the 
stellar atmospheres, which produces large element-to-element variations.

The effective temperatures for the G-jump and M-jump in each cluster
are determined by observing where the observed HB distribution
deviates from the ZAHB distribution transferred with synthetic spectra at
standard cluster composition ({\it green curves}).
Although the M-jump is well-defined in NGC~2808, it is spread over a
significant color range in NGC~6715, NGC~6388, and NGC~6441.
Nonetheless, a transition temperature of 20,000~K is consistent with
the observed M-jump in each cluster.  For NGC~2808 and NGC~6715, the
temperature of the G-jump is close to the $\sim$11,500~K temperature
reported at the time of its original discovery (Grundahl et al.\ 1998,
1999).  In contrast, the G-jump occurs at $\sim$14,000~K in NGC~6388
and $\sim$13,500~K in NGC~6441, although given the small number of
stars in the vicinity of the NGC~6441 G-jump, this value is
uncertain at the level of a few hundred degrees.

Although the models in Figure~6 employ metallicities appropriate for
each cluster, these clusters exhibit variations in He enhancement
along the HB, and the assumed $Y$ has a small but non-negligible effect on
the alignment of the models to the data (see Figure 4).  For this reason,
in Figure~7 
we also characterize the effective temperatures of the discontinuities
using ZAHB models representing sub-populations with strong He
enhancement ($\Delta Y$~=~0.17).  The temperature determinations shown
in Figures~6 and 7 bracket the possibilities in each cluster.  Even
if one tries to minimize the temperature distinctions in the G-jump for
each cluster, by assuming the coolest estimates for the metal-rich
clusters (NGC~6388 and NGC~6441) and the hottest estimates for the 
clusters at lower metallicity (NGC~2808 and NGC~6715), the metal-rich clusters
have G-jump temperatures at least 1,100~K hotter than those in the
other clusters.

\subsubsection{Caveats}

Figures~6 and 7 simulate the effects of radiative levitation in BHB stars by
assuming a super-solar metallicity between the G-jump and M-jump.  The
evidence for radiative levitation blueward of the G-jump is well
supported by spectroscopy of BHB stars (e.g., Moehler et al.\ 1999,
2000; Behr 2003; Pace et al.\ 2006).  In the CCP, the
observed HB locus returns to alignment with the standard ZAHB
distribution (i.e., with no atmospheric enhancement of metallicity)
blueward of the M-jump, but the complete cessation of radiative levitation
cannot be the origin of the M-jump; instead, some other effect is likely
counterbalancing the deviation associated with radiative levitation.
There are few metallicity measurements in globular clusters for EHB
stars (i.e., at $T_{\rm eff} \gtrsim 20,000$~K), 
although Brown et al.\ (2012) found that
some (but not all) 
of these stars in NGC~2808 exhibited super-solar Fe abundances.
The sdB stars of the Galactic field population are the analogs of the
EHB stars in globular clusters, and Geier et al.\ (2010) found that
the Fe enhancement in sdB stars hotter than 20,000~K is similar to
that in the BHB population of globular clusters.  Furthermore,
blueward of the G-jump, the enhancement of atmospheric metals is
accompanied by a corresponding depletion in atmospheric He.
In $\omega$~Cen, this He depletion continues at temperatures well past
20,000~K; in fact, the depletion continues until $T_{\rm eff} > $~32,000~K,
where the blue-hook stars exhibit atmospheres greatly enhanced in He
(Moehler et al.\ 2011).  A similar result was found in NGC~6752, with
surface He significantly depleted over the entire range of 12,000~K --
32,000~K (Moni Bidin et al.\ 2007).

Besides this evidence, we can demonstrate with our own data that the
M-jump cannot be induced by the restoration of normal atmospheric
abundances at temperatures above 20,000~K.  
In Figure~8, we show the
same data and models that appeared in Figures~6 and 7,
but now the
ordinate uses $m_{F336W}$, similar to the $U$ band employed by Grundahl
et al.\ (1999) and Momany et al.\ (2002) in their CMDs characterizing the
G-jump and M-jump.  
As noted previously, the use of [Fe/H]~=~0.5 spectra is a crude 
approximate for the abundance variations incurred through atmospheric
diffusion, and the HB exhibits a range of $Y$ values at any particular
color, due to dispersions in RGB mass loss.  It is impossible to disentangle
these complexities using broad-band photometry.
However, the purpose of
the comparison here is to demonstrate the qualitative effects of
radiative levitation in the atmosphere (and the effects of He abundance
in the stellar structure models, which will be discussed below).
The onset of radiative levitation blueward of
$m_{F275W}-m_{F438W} = -0.2$~mag causes an upward jump in the $m_{F336W}$
photometry, such that the sense of the shift is the same in the model
and the data.  However, the cessation of radiative levitation blueward
of $m_{F275W}-m_{F438W} = -1.2$~mag causes a downward shift in the model,
in contrast to the data.  Thus, the onset of radiative levitation
explains the G-jump in both the CCP (Figures~6 and 7) and the
CMD (Figure~8), but the cessation of radiative levitation does not
simultaneously reproduce the behavior of the M-jump in both diagrams.

%fig8
\begin{figure*}[t]
\begin{center}
\includegraphics[width=6.5in]{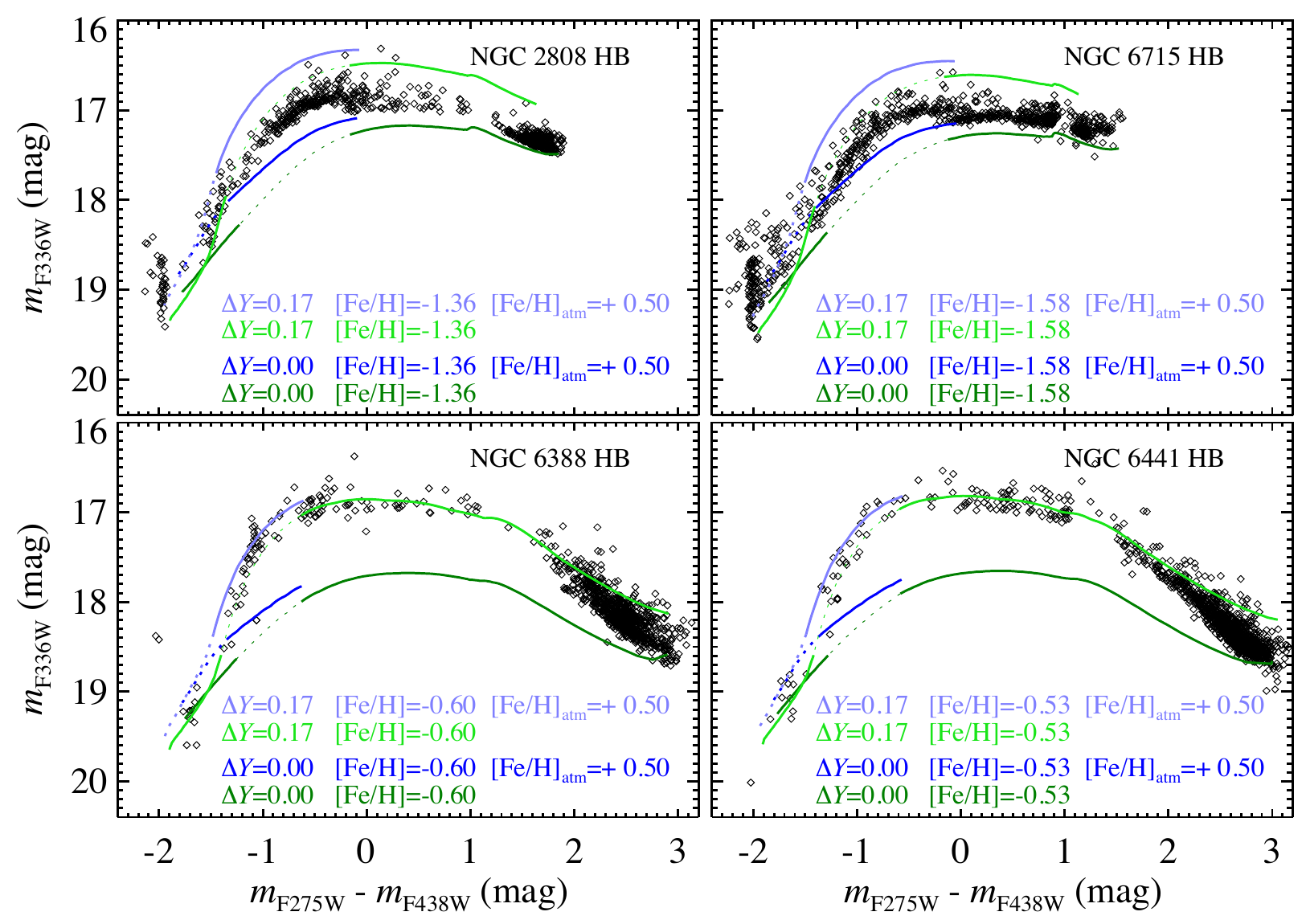}
\end{center}
\caption{The same data and models shown in Figures~6 and 7, but
  the ordinate has been replaced with $m_{F336W}$ ($U$).  Although
  the G-jump and M-jump were each discovered in CMDs of $U$ vs.\ color,
  these discontinuities are more obvious in the CCP 
  than they are in the CMD here. 
  The theoretical ZAHB distributions 
  were normalized in luminosity such that the 
  $\Delta Y = 0$ model aligns with the base of the red clump, while
  the color alignment is the same as that shown in Figures~6 and 7.
  With broad-band photometry, it is impossible
  to disentangle the effects of He enhancement 
  (accounted in the stellar structure
  models) and the radiative levitation of metals (accounted in the 
  synthetic spectra used to transfer them to the observable plane), 
  but the purpose here is to demonstrate qualitatively 
  these effects in the CMD.   The onset of radiative levitation
  for stars hotter than the G-jump causes an upward shift in both the model 
  (via synthetic spectra with super-solar abundances) and the 
  data.  The cessation of radiative levitation for stars hotter than the M-jump
  causes a downward shift in the model, in contrast
  to the data, which exhibit an upward shift.  Thus, the complete cessation
  of radiative levitation on the EHB does not explain the M-jump, even
  though it reproduces the behavior in the CCP (Figures~6 and 7).
  Redward of the G-jump, the observed
  HB luminosity in NGC~6715 is close to that of the $\Delta Y$~=~0 model, 
  indicating
  little He enhancement of its BHB stars, while the luminosity in NGC~2808
  falls between the $\Delta Y$~=~0 and $\Delta Y$~=~0.17 models, consistent with
  a moderate He enhancement.  In contrast, the BHB stars
  in NGC~6388 and NGC~6441 are much brighter than the ZAHB with
  no He enhancement ($\Delta Y$~=~0),
  but consistent with the ZAHB representing strong He enhancement
  ($\Delta Y$~=0.17). The hotter 
  G-jump in NGC~6388 and NGC~6441 (see Figures~6) is likely due to the
  fact that their BHB stars were born at greatly enhanced He abundance.
  The G-jump is more difficult to discern in this CMD than in the CCP
  (compare with Figures~6), but it is still apparent in each
  cluster.  In NGC~6388, there may be multiple deviations in the vicinity
  of the G-jump, near $m_{\rm F275W} - m_{\rm F438} \approx -0.6$~mag (also 
  evident in Figure~6) and $m_{\rm F275W} - m_{\rm F438} \approx -0.3$~mag
  (closer to the usual temperature of the G-jump).}
\end{figure*}

Although a discussion of 
the red clump morphology in these clusters is beyond the
scope of the current paper, we note some clusters exhibit red clump
distributions in the CCP (Figure~5) that are distinct
from that of NGC~2808.  The distinction can be modest (e.g., NGC~5286
and NGC~5904) or severe (e.g., NGC~6388 and NGC~6441), and is likely driven
by metallicity effects in the reddest HB stars.  However, even if
we calculate ZAHB distributions at the appropriate metallicity
for NGC~6388 and NGC~6441, there is a significant mismatch between
the models and data in Figures~6 and 7. These stars are as cool as those on
the MS, where the sensitivity of the $C_{F275W,F336W,F438W}$ index to CNO
abundances makes it a useful diagnostic in the exploration of multiple
populations (see Milone et al.\ 2012).  The distortions of the red
clump here may be related to the distinct MS morphologies of these two
metal-rich clusters.  Bellini et al.\ (2013b) demonstrated that the MS of
NGC~6441 is clearly split into two branches, while that of NGC~6388 is
broadened but not split.  They hypothesized that the second generation
of stars in each cluster has a similarly enhanced He abundance but
distinct CNO abundances.  Those CNO variations may explain why the 
two clusters exhibit distinct red clump morphologies in Figures~6 and 7.
The predicted morphology of HB evolutionary
tracks with CNO-enhanced mixtures (Pietrinferni et al.\ 2009) seems to be
consistent with this possibility (see also the optical
analysis of NGC~1851 by Gratton et al.\ 2012).

\subsubsection{Origin of the G-jump and M-jump}

The G-jump is associated with a sharp increase in atmospheric metallicity,
decrease in atmospheric He, and decrease in stellar rotation.  We
can now add that the G-jump is almost universally consistent in
effective temperature, with the notable exceptions of NGC~6388 and
NGC~6441.  The most likely reason for the increased temperature of the
G-jump in NGC~6388 and NGC~6441 is that the BHB stars in these two
metal-rich clusters are significantly enhanced in He, compared to BHB
stars in relatively metal-poor clusters; at higher metallicities,
a larger He abundance is needed to populate the BHB.  As mentioned
previously, all globular clusters appear to exhibit sub-populations
with distinct chemical compositions, but the phenomenon is strongest
in massive globular clusters, where there exist sub-populations
enhanced in He up to $Y\sim 0.4$ ($\Delta Y \sim 0.17$; 
see Piotto et al.\ 2015 and references therein).  

At a fixed age, the MS turnoff mass decreases as He increases, and
thus for a given amount of RGB mass loss, MS stars at higher $Y$ tend
to produce HB stars of lower mass and higher effective temperature.
He enhancement also affects the HB luminosity.
On the red clump and BHB, He-enhanced stars are brighter than normal, 
due to the larger energy output of their hydrogen shells that results 
from their lower envelope opacity and higher envelope mean molecular weight 
(see, e.g., Sweigart 1987).  In contrast, on the EHB, He-enhanced stars are 
fainter than normal, since they have smaller He core masses 
(see, e.g., Sweigart \& Gross 1978) and insufficient envelope masses to 
support an active hydrogen shell (Valcarce et al.\ 2012).
NGC~2808 is an
instructive example for these effects.  From the first parameter of HB
morphology (metallicity), one might expect an HB distribution that
does not extend to the end of the EHB, given its intermediate
metallicity ([Fe/H]~=~$-1.36$; Walker 1999)\footnote{The Walker et
  al.\ (1999) metallicity is on the Zinn \& West (1984) metallicity
  scale.  Recently, Carretta (2015) found [Fe/H]~=~$-1.129 \pm 0.005
  \pm 0.034$ on the UVES scale, but the distinction makes no
  difference in our analysis here.}.  However, we know that massive
clusters tend to host significant EHB populations associated with
He-enriched populations (see Milone et al.\ 2014). This is true in
NGC~2808 ($M_V = -9.4$~mag; Harris 1996), which hosts sub-populations
with varying amounts of He enhancement (up to $Y\sim 0.4$; D'Antona et
al.\ 2005; Piotto et al 2007); its hotter HB stars are generally drawn
from sub-populations with higher $Y$, with increasingly large
deviations from the luminosity of the canonical HB (e.g., D'Antona \&
Caloi 2004; Brown et al.\ 2010; Dalessandro et al.\ 2011).  In the
atmospheres of its BHB stars, Marino et al.\ (2014) also found that He
is enhanced, until the point of the Grundahl jump, where He is
depleted through gravitational settling.  The varying progenitor
populations for HB stars as a function of color is reinforced by the
work of Gratton et al.\ (2011), who found that the BHB stars in
NGC~2808 are O-poor and Na-rich (corresponding to the blue sub-population 
on the MS), while its red HB stars are O-rich and Na-poor (corresponding to the
red sub-population on the MS).  Another example of these effects can
be seen in the comparison of M3 and M13, two intermediate-metallicity
clusters with distinct HB morphology that can be traced to He enhancement
($\Delta Y \sim 0.02$--0.04) in M13 (Dalessandro et al.\ 2013).

Along similar lines, the metallicities of NGC~6388 ([Fe/H]~=~$-0.60$;
Piotto et al.\ 2002) and NGC~6441 ([Fe/H]~=~$-0.53$; Harris 1996) are
so high\footnote{Note 
that Carretta et al.\ (2009) find NGC~6388 and NGC~6441
to be at [Fe/H]~=~$-0.45 \pm 0.04$ and $-0.44 \pm 0.07$, respectively,
on the UVES scale, but as with NGC~2808, the distinction makes no difference
to our analysis here.} 
that they would normally produce an HB falling entirely in the
red clump.
However, the HB morphology of each cluster extends far to
the blue (Rich et al.\ 1997), including significant populations of
unusually-bright RR~Lyrae stars (Layden et al.\ 1999; Pritzl et
al.\ 2001, 2002, 2003; Corwin et al.\ 2006) that belong to
neither Oosterhoff class (Pritzl et al.\ 2000).  The HB of each cluster
is over-luminous blueward of the red clump, implying that the BHB
stars originate in a MS population enhanced to $Y\sim 0.4$ 
(or equivalently $\Delta Y \sim$ 0.14--0.17; Busso et
al.\ 2007; Caloi \& D'Antona 2007; D'Antona \& Caloi 2008; Brown et
al.\ 2010).  He enhancement is required to move a metal-rich HB star
blueward from the red clump to the BHB while increasing its luminosity.

To demonstrate the effects of $Y$ enhancement on HB luminosity, we
show in Figure~8 the same HB distributions of Figures~6 and 7, but using CMDs
with $m_{F336W}$ ($U$) on the ordinate. To ease comparisons between the clusters,
the observed HB distributions are again aligned to that of NGC~2808
(with the same color alignment employed in previous figures).
Here, the theoretical ZAHB
distributions (Brown et al.\ 2010) have been calculated for stars
at $\Delta Y=0.0$ ({\it dark green}; the same models in Figure~6) and
$\Delta Y=0.17$ ({\it light green}; the same models in Figure~7).  
As in Figures~6 and 7, the radiative levitation of metals is simulated
through the use of super-solar spectra ({\it blue}).
The abscissa alignment of the models to the data
is the same as that used in Figures~6 and 7, while the ordinate alignment
of the models to the data places the ZAHB model with $\Delta Y = 0.0$ 
at the base of the observed red clump (on the assumption that the faintest
red clump stars arise from a population unenhanced in He).
In NGC~6715, the G-jump
is clearly visible, but to the red of
the G-jump, the observed HB stars are closer to the model
for stars at $\Delta Y=0.0$, implying that
HB stars in the vicinity of the G-jump were born with little He
enhancement.  In NGC~2808, the G-jump is again clearly visible,
but to the red of the G-jump, there is a more
significant luminosity difference between the observations and the
model for stars at $\Delta Y=0.0$, implying that the HB stars in the
vicinity of the G-jump were born with some He enhancement.  For
reference, D'Antona \& Caloi (2004) used photometry to estimate that
these stars have moderate enhancement,
with $\Delta Y \sim$0.04--0.06
(see also Dalessandro et al.\ 2011), while Marino et
al.\ (2014) used spectroscopy to measure a He enhancement of 
$\Delta Y = 0.09 \pm 0.01 \pm 0.05$ (internal plus systematic uncertainty).  
For NGC~6388 and NGC~6441, the HB stars near the G-jump are much more
luminous than the model for $\Delta Y = 0.0$, and nearly as
bright as the model for $\Delta Y=0.17$.  This is in agreement with previous
photometric results (e.g., Busso et al.\ 2007; Caloi \& D'Antona 2007;
D'Antona \& Caloi 2008; Brown et al.\ 2010; Bellini et al.\ 2013b).  
Furthermore, the G-jump appears more complicated for these two clusters.  
For NGC~6441, the paucity of stars in the vicinity of the G-jump makes the
jump difficult to discern (at least compared to the clear deviation in
Figures~6 and 7).  For NGC~6388, however, there appears to be a pair of
jumps: one at the usual temperature for the G-jump (i.e., near
11,500~K), and a hotter one (i.e., near 14,000~K, corresponding to the
obvious deviation in Figures~6 and 7); 
these appear as excursions from the dominant stellar locus.

If the BHB stars of NGC~6388 and NGC~6441 were born with a large He
enhancement, can this explain the increased temperature for the G-jump
in these clusters?  To investigate this point, we show in Figure~9 the
behavior of the convective zones due to H, \ion{He}{1}, and
\ion{He}{2} ionization near the surface of HB stars, as a function of
effective temperature, using evolutionary models calculated at 
metallicities appropriate for NGC~6388 
and NGC~2808.  The figure shows the convective
zones for HB stars with $\Delta Y=0.03$ ({\it red shading}) and
$\Delta Y=0.17$ ({\it blue shading}).  Sweigart (2002) and Cassisi \&
Salaris (2013) used similar figures (with no He enhancement) to
demonstrate how the surface convection from the \ion{He}{1} ionization
zone disappears at temperatures hotter than $\sim$12,000~K, enabling
the onset of radiative levitation in BHB stars and the appearance of
the G-jump.  Here, we show the convection zones for two different $Y$
assumptions, demonstrating that the \ion{He}{1} convection is
sensitive to initial $Y$, while the \ion{He}{2} and H convection zones are not.
Comparing the convective zones in the top 
(intermediate metallicity) and bottom (high metallicity) panels, it 
is clear that $Y$ (and not [Fe/H]) is the critical parameter governing
the behavior of the \ion{He}{1} convection zone, and thus the G-jump, although
we note that there are other changes in the convection zones that depend
upon [Fe/H]. Specifically, increasing [Fe/H] shifts the termination of the H
convection zone hotter by 500--1000~K (depending upon $Y$) and extends
the tail of the \ion{He}{2} convection zone to both hotter temperatures and
shallower depths.

%fig9
\begin{figure}[t]
\includegraphics[width=3.4in]{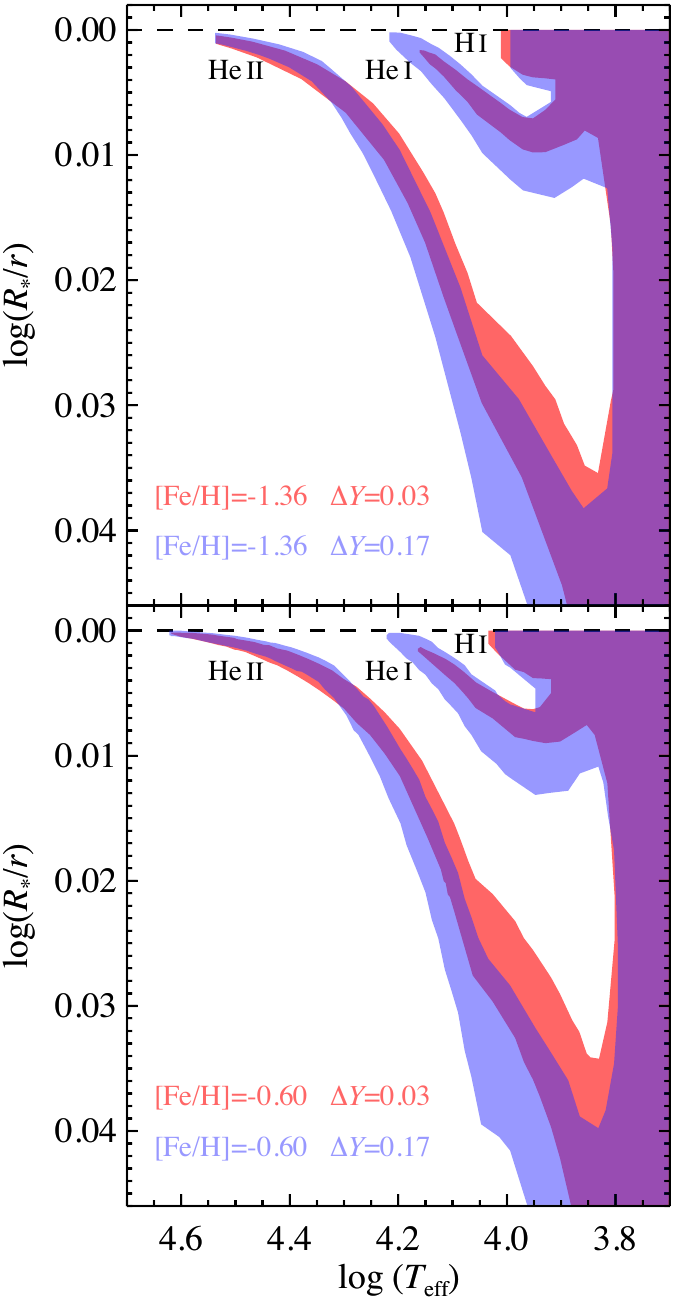}
\caption{The location of the convection zones ({\it shading}) in ZAHB stars
as a function of effective temperature, relative to the stellar surface
({\it dashed line}), assuming the metallicity of
NGC~2808 ({\it top panel}) and NGC~6388 ({\it bottom panel}).  
When the He abundance is increased from $\Delta Y=0.03$ ({\it red shading}) to
$\Delta Y=0.17$ ({\it blue shading}), 
the transition in the \ion{He}{1} convective 
zone shifts to higher effective temperature by $\sim$2,000~K, and the 
\ion{He}{1} convective zone moves closer to the surface.
If the G-jump is normally associated with the \ion{He}{1} convective zone,
this temperature shift at high $Y$ may explain the bluer G-jump in 
NGC~6388 and NGC~6441.
If the G-jump is normally associated with the H convective zone, the surface 
encroachment of the \ion{He}{1} convective zone at high $Y$ may also 
explain the bluer G-jump in NGC~6388 and NGC~6441.  Note that evolutionary
effects, turbulence, and mass loss complicate this interpretation;
these models only serve to provide possible explanations for the 
observed G-jump behavior in a qualitative sense.  Similarly,
if the M-jump is associated with the \ion{He}{2} convective zone, the 
observed consistency of the M-jump temperature (even in NGC~6388 and NGC~6441)
might be due to the insensitivity of the \ion{He}{2} convective zone to $Y$.
}
\end{figure}

With this behavior in mind, the convection zones in the vicinity of
the G-jump (from H and from \ion{He}{1}) offer two possible
explanations for the G-jump.  Qualitatively, a convective zone acts as
a fully mixed reservoir of matter having the original chemical
composition, which minimizes the effects of diffusion (gravitational
settling and radiative levitation).  The G-jump could be due to the
\ion{He}{1} ionization, as hypothesized by Sweigart (2002) and Cassisi
\& Salaris (2013).  In this view, the BHB stars in most clusters are
born with little to modest He enhancement, such that there is little
variation in the effective temperature of the G-jump.  In NGC~6388 and
NGC~6441, the BHB stars have a significant enhancement near 
$\Delta Y\sim 0.17$; along the HB, this
shifts the cessation of \ion{He}{1} ionization to hotter effective
temperatures by $\sim$2,000~K -- the same shift observed.  In this
scenario, it is unclear if the G-jump in NGC~2808 (with stars 
near $\Delta Y\sim 0.09$) would be so similar to the G-jump in less massive
clusters that exhibit little He enhancement in their
sub-populations (see Figure~5).  Alternatively, 
the G-jump in most clusters could be due to the H
convection zone.  In this view, the G-jump remains at constant
effective temperature in most clusters because the H convection is
insensitive to $Y$.  However, for BHB stars with He abundances 
near $\Delta Y\sim 0.17$, the
\ion{He}{1} ionization shifts to hotter effective temperature and
moves closer to the stellar surface.  This would make the G-jump
appear at hotter effective temperature (as in Figure~5) if it arises
from \ion{He}{1} convection, or even multiple temperatures (as might
be implied by Figure~8) if it arises from both H and \ion{He}{1}
convection.  With both of these scenarios, the temperatures observed
for the G-jump do not exactly coincide with the transitions in the
modeled convection zones, even if the behavior is qualitatively
consistent.  It is likely the case that other parameters, such as
turbulence, play a role in the exact location of the jumps;
the region that is mixed at the surface may not coincide with
the formal convective boundary.  Along
these lines, we note that Michaud et al.\ (2011) reproduce the G-jump
at 11,500~K by invoking a fully mixed region near the surface of the
star, with a mass 10$^{-7}$~$M_\odot$, possibly driven by turbulence
or mass loss.

Another complication concerns the direction of the HB 
evolution beyond the ZAHB.  As known
since the work of Sweigart \& Gross (1976), 
BHB stars with normal He abundances can slowly evolve redward 
toward cooler effective temperatures, especially at low to intermediate
metallicities.  However, when their He enhancement
is $\Delta Y=0.17$, the HB stars in the vicinity of the G-jump 
can rapidly evolve blueward
from temperatures cooler than 11,500~K to temperatures near 15,000~K.
Depending upon the relative timescales of the various factors at work
(surface convection, turbulence, mass loss, radiative levitation, gravitational
settling), the distinct evolutionary paths for high-$Y$
stars may also push the G-jump toward higher effective temperatures in
clusters like NGC~6388 and NGC~6441. An exploration of these
effects is currently underway (Tailo et al., in prep.).

As noted previously, the M-jump cannot be due to a simple disruption
of radiative levitation at temperatures hotter than 20,000~K.
Looking at Figure~9, it is worth noting that the \ion{He}{2}
convection zone begins to encroach upon the surface near this
temperature, and the behavior is independent of He abundance.
It may be a coincidence that the M-jump is also independent of 
He abundance, even in the two clusters exhibiting a hotter G-jump
(NGC~6388 and NGC~6441), but we speculate that the M-jump may be associated
with the \ion{He}{2} convection zone.

\section{Summary}

Using UV and blue photometry for 53 Galactic globular clusters, we
have shown that the discontinuities in their HB distributions are
remarkably consistent. Globular clusters are now known to host complex
populations with variations in chemical composition, but these HB
discontinuities reflect universal transitions in atmospheric
phenomena, and not abundance distinctions in their MS progenitors.
That said, the effective temperature for one of these discontinuities,
the G-jump, is $\sim$1,000--2,000~K hotter in NGC~6388 and NGC~6441.  This
shift is likely due to the fact that these two clusters host BHB stars
greatly enhanced in He ($\Delta Y \sim 0.17$), which affects the behavior of
the \ion{He}{1} convective zone and its role in disrupting radiative
levitation.

Although the complexity of globular cluster populations was originally
recognized in the most massive globular clusters, its ubiquity became
more apparent with appropriate photometry.  The history of blue-hook
stars is following a similar path.  This is not because massive
clusters provide more chances to find a star following a relatively
rare evolutionary avenue; instead, these clusters host sub-populations
significantly enhanced in He, which leads to a hotter HB morphology.
While blue-hook stars comprise a tiny fraction of the population in
any globular cluster, we have shown that these products of extreme
mass loss can be found in most of the bright globular clusters.  The 
census of clusters hosting blue-hook stars has increased to 23, with
nearly all of them residing in the brighter half of the Galaxy's 
globular cluster system.

\acknowledgements

Support for program GO-13297 was provided by NASA through a grant from
the Space Telescope Science Institute, which is operated by the
Association of Universities for Research in Astronomy, Inc., under
NASA contract NAS 5-26555.  
S.C.\ and G.P.\ recognize partial support
by the IAC (grant P301031) and the Ministry of Competitiveness
and Innovation of Spain (grant AYA2010-16717); S.C., G.P., and
F.D.\ recognize partial support
by PRIN-INAF 2014 (PI: S.\ Cassisi).
S.O.\ and G.P.\ also acknowledge partial support by the
Universit\`a degli Studi di Padova Progetto di Ateneo CPDA141214
``Towards understanding complex star formation in Galactic globular
clusters.''  A.P.M.\ acknowledges support by the
Australian Research Council through Discovery Project grant
DP120100475.
F.R.F.\ and E.D.\ acknowledge the support from the Cosmic-Lab
project (web site: http://www.cosmic-lab.eu) funded by the
European Research Council, under contract ERC-2010-AdG-
267675. A.P.\ acknowledges support from PRIN-INAF2012 (PI: L.\ Bedin).
S.O.\ gives thanks for the support of the University of Padova.

\end{document}